\documentclass[preprint,12pt,authoryear,review]{elsarticle}
\usepackage[utf8]{inputenc}
\usepackage{amsmath, amsthm, amsfonts, amssymb, amscd}
\usepackage{graphicx}
\usepackage{enumerate}
\usepackage{natbib}
\usepackage{url} 
\usepackage{xcolor}
\usepackage{graphicx}
\usepackage[linesnumbered,ruled]{algorithm2e}
\usepackage{booktabs}
\usepackage{subcaption}
\usepackage{tikz}
\usepackage{datatool}
\usepackage{etoolbox}
\usepackage{float}

\DTLnewdb{bibnotes}

\def\bibnote#1#2{%
  \DTLnewrow{bibnotes}
  \DTLnewdbentry{bibnotes}{mylabel}{#1}
  \DTLnewdbentry{bibnotes}{mynote}{#2}
}

\makeatletter
\patchcmd{\@lbibitem}%
  {\item[\hfil\NAT@anchor{#2}{\NAT@num}]}%
  {%
    \item[\hfil\NAT@anchor{#2}{\NAT@num}]%
    \DTLforeach[\DTLiseq{\mylabel}{#2}]{bibnotes}{\mylabel=mylabel,\mynote=mynote}{{\mynote}}
  }{}{\message{^^JPatching failed^^J}}%

\newcommand{\sqdiamond}[1][fill=black]{\tikz [x=1.2ex,y=1.85ex,line width=.1ex,line join=round, yshift=-0.285ex] \draw  [#1]  (0,.5) -- (.5,1) -- (1,.5) -- (.5,0) -- (0,.5) -- cycle;}%
\newcommand{\MyDiamond}[1][fill=black]{\mathop{\raisebox{-0.275ex}{$\sqdiamond[#1]$}}}

\journal{Spatial Statistics}

\begin{document}

\begin{frontmatter}



\title{Flexible Spatio-Temporal Hawkes Process Models for Earthquake Occurrences}


\author[label1]{Junhyeon Kwon}
\author[label2]{Yingcai Zheng}
\author[label1]{Mikyoung Jun\corref{cor1}}
\ead{mjun@central.uh.edu}

\cortext[cor1]{Corresponding author}

\affiliation[label1]{organization={Department of Mathematics, University of Houston},
            country={United States of America}}
            
\affiliation[label2]{organization={Department of Earth and Atmospheric Sciences, University of Houston},
            country={United States of America}}
            
\begin{abstract}
Hawkes process is one of the most commonly used models for investigating the self-exciting nature of earthquake occurrences. However, seismicity patterns have complicated characteristics due to heterogeneous geology and stresses, for which existing methods with Hawkes process cannot fully capture. This study introduces novel nonparametric Hawkes process models that are flexible in three distinct ways. First, we incorporate the spatial inhomogeneity of the self-excitation earthquake productivity. Second, we consider the anisotropy in aftershock occurrences. Third, we reflect the space-time interactions between aftershocks with a non-separable spatio-temporal triggering structure. For model estimation, we extend the model-independent stochastic declustering (MISD) algorithm and suggest substituting its histogram-based estimators with kernel methods. We demonstrate the utility of the proposed methods by applying them to the seismicity data in regions with active seismic activities.
\end{abstract}


\begin{highlights}
\item Proposed Hawkes process model describes aftershock occurrences flexibly.
\item Proposed model accounts for spatially varying aftershock productivity.
\item Proposed model allows space-time interaction and spatial anisotropy of aftershock occurrences.
\item Proposed model improves the forecast accuracy.
\item This paper presents the changes in seismicity after large magnitude earthquakes.
\end{highlights}

\begin{keyword}
Anisotropy \sep Earthquake \sep Hawkes Process \sep Inhomogeneous model \sep Spatio-Temporal Nonseparability \sep Spatio-Temporal Point Process

\end{keyword}

\end{frontmatter}

\section{Introduction}
\label{sec:intro}
Earthquakes are well-known phenomena in nature with self-exciting properties in space and time \citep{vanderelst2010}. The stress changes by an earthquake can cause and trigger additional earthquakes in the nearby region. These ``triggered" earthquakes can then trigger more earthquakes, and this process leads to space-time clusters with a branching structure. In this paper, we refer to a triggering event and its triggered events as the `mainshock' and `aftershocks,' respectively. It is worth noting that in this definition a small-magnitude earthquake can trigger a larger-magnitude aftershock.

\begin{figure}[!hbt]
    \centering
    \includegraphics[width=0.7\textwidth]{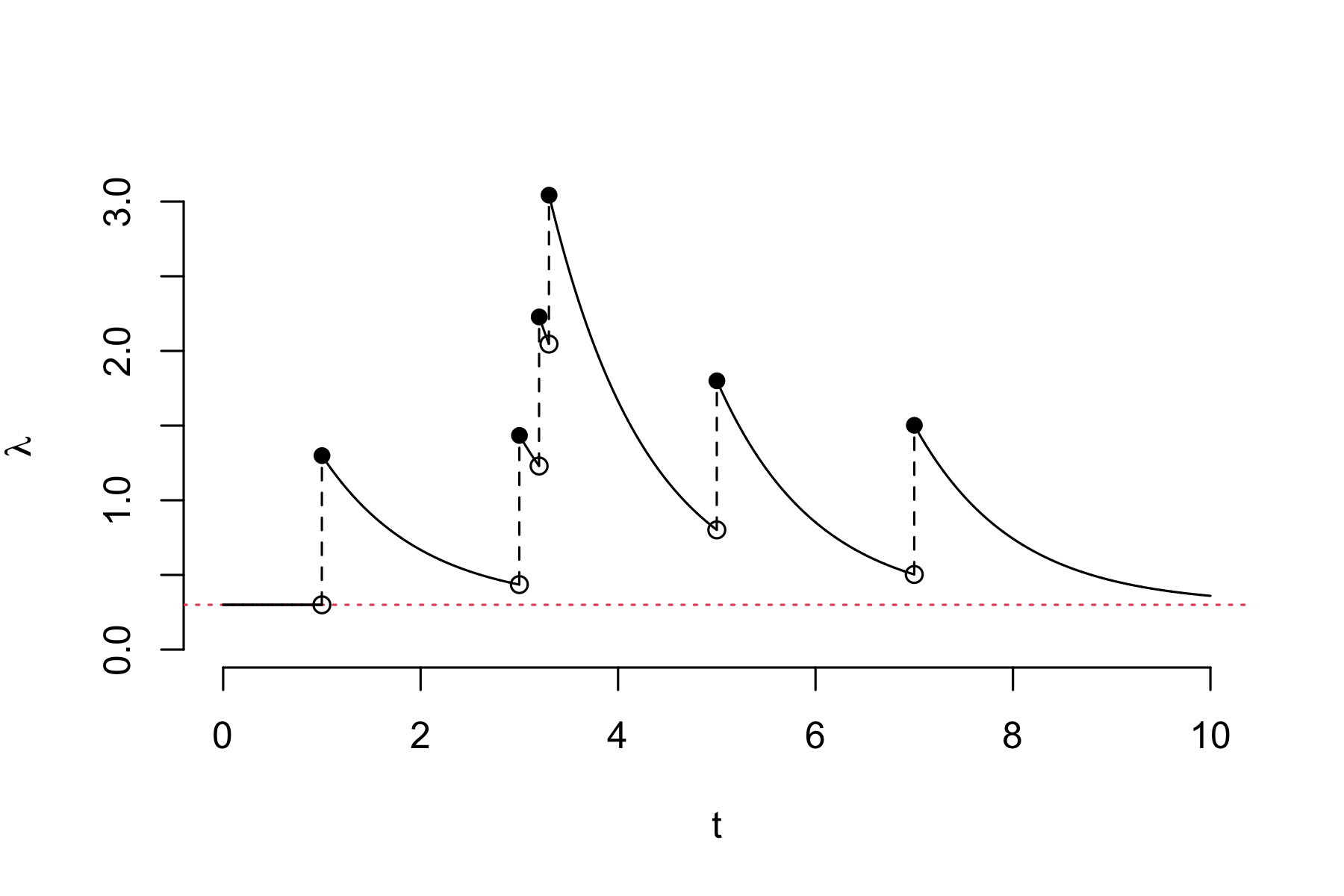}
    \caption{An example of conditional occurrence rate for a temporal Hawkes process. Dotted red line represents the background rate $\mu$ (assumed to be constant in this example).}
    \label{fig:temporal_hawkes}
\end{figure}

This paper models the earthquake occurrences with the {\it Hawkes process} models that are useful to explore data with self-exciting properties. One simple form of the Hawkes processes is their temporal version. Assume that we observe events at time points $t_1, \, t_2, \, t_3, \, \cdots $ that have self-exciting properties.  Conditional on the past events up to time $t$, $\mathcal{H}_t$,  it models the event occurrence rate as
\begin{equation}\label{eq:hawkes_temporal}
    \lambda(t|\mathcal{H}_t) = \mu(t) + \sum_{j: t_j < t} g(t - t_j),
\end{equation}
where $\mu(t) \geq 0$ is the so-called background rate from the background process. The temporal triggering function, $g(\Delta t) $, is a function of the time lag $\Delta t > 0$. A commonly used form of $g$ is $a\cdot\exp(-\Delta t/b)$ for positive constants $a$ and $b$. Figure \ref{fig:temporal_hawkes} illustrates the changes in $\lambda(t)$ in the time interval $[0,10]$ as the events occur at $t = 1, \, 3, \, 3.2, \, 3.3, \, 5, \, 7$, with $a=b=1$. Events become more likely to occur as the triggering effects of the previous events accumulates. We can also observe that the triggering effect is diminishing as time passes due to the structure assumed for $g$.

Hawkes processes have a wide range of applications including crime occurrences \citep{mohler2011self, zhu2022spatiotemporal}, terrorism \citep{jun2022flexible}, social media \citep{yuan2019multivariate}, and infectious disease such as COVID-19 \citep{browning2021simple}. For earthquakes applications, \cite{ogata1988statistical} suggested the Epidemic-Type Aftershock Sequence (ETAS) model, which is a temporal Hawkes process for earthquake occurrences. It assumes that the earthquake productivity can be attributed to the background process for the mainshocks plus the aftershocks triggered by previously occurred events. Later, \cite{ogata1998space} suggested a spatio-temporal ETAS model. It models the mainshocks by the spatio-temporal Poisson point process, and the triggering effect by $\kappa(m)g(\Delta x, \Delta y, \Delta t)$. Here, $\kappa(m)$ is the aftershock productivity (the expected number of the triggered aftershocks) for the earthquake with magnitude $m$, and $g(\Delta x, \Delta y, \Delta t)$ is a spatio-temporal triggering density with respect to the lags in longitude, latitude, and time.

\begin{figure}[h!]
     \centering
     \begin{subfigure}[b]{0.35\textwidth}
         \centering
         \includegraphics[width=\textwidth]{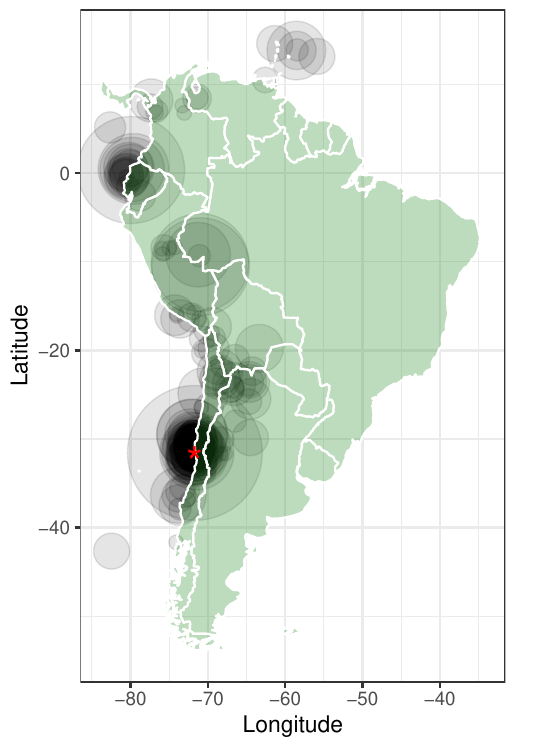}
         \caption{Spatial pattern}
         \label{fig:sam_2015a}
     \end{subfigure}
     \hfill
     \begin{subfigure}[b]{0.61\textwidth}
         \centering
         \includegraphics[width=\textwidth]{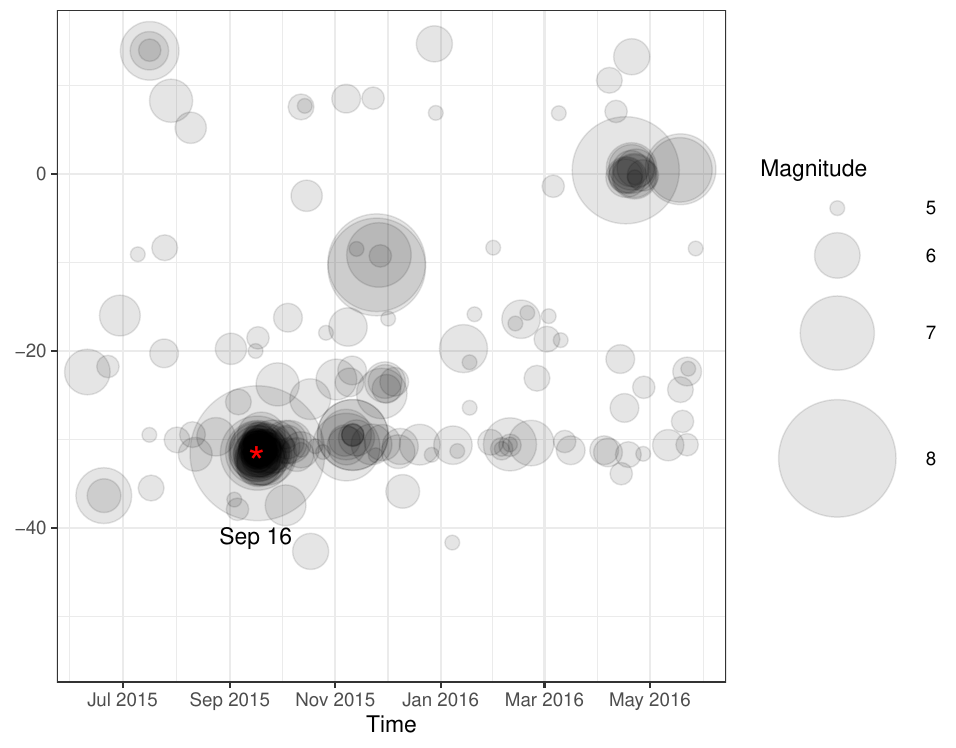}
         \caption{Spatio-temporal pattern}
         \label{fig:sam_2015b}
     \end{subfigure}
     \caption{Clustered structure of earthquake occurrences $(M \geq 5.0)$ in South America between June 2015 and May 2016. (a) Spatial distribution; (b) Spatio-temporal distribution. The mainshock epicenter is marked by an asterisk in both figures and its date is noted in the figure b.}
     \label{fig:sam_2015}
\end{figure}

It is helpful to visualize the earthquake occurrences to build a model that reflects the nature of earthquake occurrences. 
Figure \ref{fig:sam_2015} illustrates the spatial and spatio-temporal patterns of the earthquake occurrences of magnitude 5.0 or greater in South America during a one-year period, beginning in June 2015. The asterisk symbol represents the epicenter and occurrence time of the magnitude 8.3 earthquake that struck Chile on September 16, 2015. We can observe a distinct spatial cluster of earthquakes around the mainshock in Figure \ref{fig:sam_2015a}, and spatio-temporal cluster in Figure \ref{fig:sam_2015b}. These clusters suggest that a large earthquake in September 2015 triggers more aftershocks compared to the occurrences of smaller earthquakes. Furthermore, the number of triggered aftershocks appears to vary depending on both the location and the magnitude. For example, there was an earthquake with magnitude 7.6 at latitude around $L = -10^\circ$ in November 2015, and another with magnitude 7.8 at latitude around $L = 0^\circ$ in April 2016. Considering the small difference in their magnitudes, these two events in November 2015 have much fewer subsequent earthquakes compared to the one in April 2016. In addition to the aftershock productivity and its resulting cluster structure, earthquake activity can be seen mainly on the western coast of South America, known for its active tectonic plate subduction. This example illustrates the need for a flexible Hawkes process model with spatially inhomogeneous background rate, space-time triggering density whose decay rate accounts for complex cluster structure, and aftershock productivity which depends both on the magnitude and location of the mainshock.

We consider {\it nonparametric} ETAS models, as it may be restrictive to assume a certain parametric form for a complex process such as earthquake occurrences. Existing works on nonparametric approaches \citep{marsan2008extending, fox2016spatially, gordon2021nonparametric} used  histogram estimators to describe the triggering function. For example, a triggering function $g(\Delta t)$ in \eqref{eq:hawkes_temporal} is estimated in the form
\begin{equation*}
    \hat{g}(\Delta t) = \sum_{b=1}^B \gamma_b I_{[\tau_{b}, \tau_{b+1})}(\Delta t),
\end{equation*}
where the temporal lag is partitioned into $B$ bins with edge points: $\tau_1<\tau_2<\cdots<\tau_{B+1}$, $\gamma_b>0$ is the height of the histogram in the bin $[\tau_b, \tau_{b+1})$, and $I$ is an indicator function such that $I_{[\tau_b, \tau_{b+1})}(\Delta t) = 1$ if $\Delta t \in [\tau_b, \tau_{b+1})$ and 0 otherwise. However, these existing nonparametric ETAS models have limited flexibility due to the assumption of a locally constant estimator. Furthermore, these methods may not be flexible enough to deal with location dependence and space-time interactions of aftershocks.

In this paper, we propose a new class of kernel-based nonparametric ETAS models to study aftershock dynamics, focusing on three aspects. First, we allow the model to have different aftershock productivity depending on the spatial location of each event as well as its magnitude. Second, to account for the anisotropy in the aftershock distribution, we use the Mahalanobis distance to reflect geological characteristics such as fault direction in the spatial domain. Third, we estimate the triggering density function for the triggering dynamics in a nonseparable manner to explain the possible space-time interaction dynamics.

Throughout the paper, we use the earthquake data from Advanced National Seismic System (ANSS) Comprehensive Catalog (ComCat) \nocite{us2017advanced}, and this can be accessed by the US Geological Survey website (\url{https://earthquake.usgs.gov/earthquakes/search/}). The plate boundary information was downloaded from GitHub repository of \nocite{HugoAhlenius} Hugo Ahlenius (\url{https://github.com/fraxen/tectonicplates}) in the GeoJSON format, which is an enhanced conversion of the data originated from \cite{bird2003updated}.

The rest of the paper is organized in the following way. Section \ref{sec:2} reviews the spatio-temporal ETAS models both for parametric and nonparametric approaches. In Section \ref{sec:3}, we propose new kernel-based ETAS models with flexible triggering functions. Earthquake data is analyzed for multiple regions and time periods in Section \ref{sec:4}. Various models are compared based on forecast accuracy, and changes in mainshock activity are investigated before and after some major earthquakes. Section \ref{sec:5} summarizes the proposed model's contributions and discusses the possibility of further extension as future research topics.

\section{Background}\label{sec:2}
In this section, we review existing spatio-temporal ETAS models, both parametric and nonparametric methods.

\subsection{Parametric ETAS models}
Let $(x,y)$ denote the longitude and latitude of the location of earthquake (epicenter), $t$ denote the time of occurrence, and $m$ denote the earthquake magnitude. Here we use the moment magnitude which is defined as a continuous value $M_w = (1.5)^{-1} \log_{10} M_0 - 6.07$ for a seismic moment $M_0$ in N$\cdot$m \citep{kanamori2004physics}. We then consider the collection of $N$ earthquake occurrences sorted in time
$$\{(x_j,y_j,t_j,m_j):\ (x_j, y_j) \in D, \ t_j - t_{j-1} \geq 0, \ t_0 = 0, \ t_N \leq T, \ 1 \leq j \leq N\}$$ on a spatial domain $D$ over a period of length $T$, in the unit of days.
The earthquake occurrences can be modeled as a spatio-temporal point process. It is described by the (first-order) intensity function
\begin{equation}\label{eq:intensity}
        \lambda(x,y,t,m) = \lim_{\Delta \mathbf{s}, \Delta t, \Delta m \rightarrow 0} \frac{E \left[N \{ B(\mathbf{s}, \Delta \mathbf{s}) \times [t, t + \Delta t) \times [m, m + \Delta m) \} \right]}{ |B(\mathbf{s}, \Delta \mathbf{s})| \Delta t \Delta m},
\end{equation}
 where $N(\cdot)$ is a counting measure of events, $B(\mathbf{s}, \Delta \mathbf{s})$ is a ball centered at $\mathbf{s} = (x,y)$ with radius $\Delta \mathbf{s}$. We write a conditional intensity $\lambda(x,y,t,m|\mathcal{H}_t)$ by replacing the numerator of \eqref{eq:intensity} with the expectation conditional on the history $\mathcal{H}_t = \{(x_j, y_j, t_j, m_j): t_j < t\}$, and this can be used to define a spatio-temporal point process \citep{diggle2013statistical}. 

In the spatio-temporal ETAS models, earthquake magnitudes are commonly considered in a separable manner for the conditional intensity as
\begin{equation*}
     \lambda(x,y,t,m|\mathcal{H}_t) = J(m) \lambda_0(x,y,t|\mathcal{H}_t),
\end{equation*}
where $J(m)$ is a density of earthquake magnitudes independent from the past events, and $\lambda_0(x,y,t|\mathcal{H}_t)$ is a conditional intensity only for location and time \citep{ogata1998space,zhuang2002stochastic, marsan2008extending,fox2016spatially}. In our study, we similarly regard the magnitude component as separable as above and focus on estimating the remaining part. For simplicity, we drop the subscript in $\lambda_0$ for the rest of this paper. As in the introduction of spatio-temporal ETAS models by \cite{ogata1998space}, the reduced conditional intensity function is commonly written as 
\begin{equation*}\label{eq:intensity_etas_original}
    \lambda(x,y,t|\mathcal{H}_t) = \mu(x,y) + \sum_{\{j:t_j < t\}} \nu(x-x_j,y-y_j,t-t_j;m_j)
\end{equation*}
where $\mu(x,y)$ is a background rate for the mainshock at the location $(x,y)$, and $\nu(x-x_j, y-y_j, t-t_j;m_j)$ represents the triggering effect at location $(x,y)$ and time $t$ from the event occurred at epicenter $(x_j,y_j)$ and time point $t_j$ with magnitude $m_j$. The triggering function $\nu$ is further divided as
\begin{equation*}\label{eq:triggering_etas_original}
    \nu(x-x_j, y-y_j, t-t_j;m_j) = \kappa(m_j) g(x - x_j, y - y_j, t - t_j;m_j),
\end{equation*}
where $\kappa(m_j)$ is the number of aftershocks that would be triggered on average by an event of magnitude $m_j$, and $g(x - x_j,y - y_j,t - t_j;m_j)$ is a density function which explains how the aftershocks of the $j$-th event would be scattered both spatially and temporally centered on the epicenter $(x_j,y_j)$ and the time of occurrence $t_j$. Spatial lags $x - x_j$ and $y - y_j$ in the triggering density are sometimes scaled based on the magnitude $m_j$ because a large-magnitude earthquake triggers aftershocks in a wider area \citep{Utsu1955ARB, utsu1970aftershocks}. As time passes and as it becomes farther away from the epicenter, the triggering function decays to 0 and the conditional intensity $\lambda(x,y,t|\mathcal{H}_t)$ converges to the background rate $\mu(x,y)$.

Parametric ETAS models \citep{ogata1998space, zhuang2002stochastic, veen2008estimation, ogata2011significant} use the results from the empirical study or make physical hypotheses to assume specific mathematical forms for $\kappa$ and $g$. The aftershock productivity function $\kappa(m)$ is commonly assumed to be in an exponential form as
$\kappa(m) = a_0 \exp(a \cdot m),$ for positive constants $a_0$ and $a$. The triggering density is usually separated into spatial and temporal components as
\begin{equation}\label{eq:trig_separable}
    g(\Delta x,\Delta y,\Delta t;m) = g_1(\Delta x,\Delta y;m) g_2(\Delta t),
\end{equation}
where $g_1$ explains the aftershock occurrences spatially with respect to the lag of longitude $\Delta x$ and the lag of latitude $\Delta y$, and $g_2$ explains the aftershocks occurrences temporally with respect to the temporal lag $\Delta t$. The temporal triggering function $g_2$ is usually assumed to follow the modified Omori formula, 
$    g_2(\Delta t) \propto \displaystyle  (1 + \Delta t/c)^{-p}$, with $p>1$ a decay rate and $c$ a positive constant \citep{utsu1957magnitudes, utsu1995centenary}. Spatial triggering function $g_1(\Delta x, \Delta y;m)$ takes various forms depending on the decay rate or the scaling of spatial lag. In \cite{ogata1998space}, two widely used decay rates were considered. The first one is Gaussian ($ \Delta \mathbf{s} = \sqrt{\Delta x^2 + \Delta y^2}$), 
$$g_1(\Delta x,\Delta y;m) \propto \exp\left(-\frac{1}{2d} \frac{\Delta \mathbf{s}^2 }{\sigma(m)}\right),$$
and the other is the inverse power law
$$g_1(\Delta x,\Delta y;m) \propto \left(1 + \frac{\Delta \mathbf{s}^2 }{d\sigma(m)}\right)^{-q},$$
where $d>0$ and $q>1$ are the parameters to be estimated, and $\sigma(m)$ is a spatial lag scaling factor that is either $\sigma(m) = 1$ or $\sigma(m) = \exp(\beta m)$ for a constant $\beta>0$.

For the estimation of $\mu$ and other parameters in $\kappa$, $g_1$, and $g_2$, one can maximize the log-likelihood 
\begin{equation*}
    \ell(\Theta) = \sum_{i=1}^N \log \lambda(x_i, y_i, t_i | \mathcal{H}_{t_i}) - \int_0^T \int \int_D \lambda(x, y, t | \mathcal{H}_t) dx dy dt
\end{equation*}
for the set of parameters $\Theta = \{\mu_1, \mu_2, \cdots, \mu_k, a_0, a, \beta, d, q, c, p\}$ \citep{daley2003introduction, reinhart2018review}. Here, $(\mu_1, \mu_2, \cdots, \mu_k)$ are the heights of a 2-dimensional histogram or the coefficients of splines.  One of the most well-known attempts is the so-called stochastic declustering method in \cite{zhuang2002stochastic}. By assigning the probability for an earthquake being a mainshock, it stochastically splits the entire earthquake population into mainshocks and aftershocks. Initially, it assumes that $\mu(x,y)$ is constant and maximizes the likelihood to estimate $\kappa(m)$, $g_1(\Delta x, \Delta y;m)$, and $g_2(\Delta t)$. Given these estimates, the probability that the $i$-th event is a mainshock can be calculated by $\rho_i = \mu(x_i, y_i) / \lambda(x_i, y_i, t_i | \mathcal{H}_{t_i})$. Then one can update the estimate of the background rate by $\hat{\mu}(x,y) = T^{-1} \sum_{i=1}^N \rho_i G(x - x_i, y - y_i),$ where $G$ is a (Gaussian) kernel with an appropriate bandwidth. Now the stochastic declustering method iterates between the estimation of ($\kappa$, $g_1$, $g_2$) and $\mu$ until convergence. \cite{veen2008estimation} proposed an EM-type algorithm that also estimates the ETAS models using the stochastic branching structure. This framework is general in that it is also used to estimate the nonparametric ETAS models from which our method is derived. As a result, we suspend the algorithm explanation for the time being and discuss it later.

The underlying mechanism of earthquake occurrences may vary from location to location due to factors we could not account for in the model. Hence, a natural extension of ETAS models would be able to incorporate location dependence property. \cite{ogata2004space} suggested a penalized likelihood estimation method of parametric model that every earthquake has its own parameters. He interpolated the value of each parameter based on Delaunay triangulation tessellated by the epicenters. \cite{harte2014etas} proposed a model which used space-time closeness between the events to allow the parameters to vary both spatially and temporally. \textcolor{black}{\cite{zhuang2015weighted} proposed weighted likelihood estimators based on residual analysis to estimate the spatially varying parameters in ETAS models.}

Another extension considered in our work is the anisotropy in the spatial pattern of the aftershocks. Earthquakes occur as relative slip on pre-existing fault planes. We use strike and dip angles to describe the fault plane orientation. The slip angle describes the relative movement on the fault plane, during an earthquake rupture, between the two blocks. The strike measures the direction of the intersection line between the Earth's surface and the fault plane, and the dip is the angle between the fault plane and the surface. As a result, we expect the epicenters of earthquakes around the same fault to be scattered in an elliptic shape, with eccentricity determined by the strike, dip, and slip. For this reason, since the introduction of space-time ETAS models, many previous works have made efforts to reflect the shape of the aftershock pattern better. \cite{ogata1998space} suggested finding a centroid of aftershock epicenters by magnitude-based clustering and fitting a bivariate normal density to define the Mahalanobis distance for the spatial lags between the events. \cite{hainzl2008impact} pointed out that considering earthquakes to have point sources can lead to overestimation of aftershock occurrences. They instead assumed that earthquakes have line sources by using rupture geometry. In a similar context, \cite{guo2015improved} accounted for anisotropy by overlapping the circular triggering density.

\subsection{Nonparametric ETAS models}

Although there have been many efforts with the parametric forms, the physical mechanism behind the earthquake occurrences is still not well understood, and the state of the rock stress is uncertain too. Therefore, nonparametric modeling can be a good alternative. A noteworthy example of such is the work by \cite{marsan2008extending}. They suggested a model-independent stochastic declustering (MISD) method which assumed a constant background rate $\mu(x,y) = \mu$. \cite{fox2016spatially} extended the method by allowing the background rate to vary  spatially. Both \cite{marsan2008extending} and \cite{fox2016spatially} assumed that the triggering function is space-time separable and did not use a spatial lag scaling factor. Each of the functions $\kappa$, $g_1$, and $g_2$ does not assume a specific model except that it has the shape of a histogram. The function's domain is partitioned into multiple bins, and MISD method estimates the histogram heights of these bins.

To determine the histogram heights of the bins, we need to introduce indicating variables
\begin{equation*}
    \chi_{ij} = \begin{cases}
    1, \text{ if the $i$-th event was triggered by the $j$-th event}\\
    0, \text{ otherwise,}
    \end{cases}
\end{equation*}
for $1 \leq i, j \leq N$. Note that $\chi_{ii} = 1$ implies that the $i$-th event is a mainshock because it triggered itself, and $\chi_{ij} = 0$ for $i<j$ because an event cannot affect the past. If we assume that all these indicating variables can be observed, the complete log-likelihood of the ETAS model becomes
\begin{equation*}\label{eq:loglik_complete}
\begin{split}
    \ell_c(\Theta) =& \sum_{i=1}^N \chi_{ii} \log \mu(x_i,y_i) + \sum_{i=1}^N\sum_{j=1}^N \chi_{ij} \log \nu(x_i - x_j, y_i - y_j, t_i - t_j;m_j)\\
    &- \int_0^T\int\int_D \mu(x,y) dxdydt\\
    &- \sum_{j=1}^N \int_0^T\int\int_D \nu(x - x_j, y - y_j, t - t_j;m_j) dxdydt,
\end{split}
\end{equation*}
where $\Theta$ consists of the heights of the bins in the histograms for $\mu$, $\kappa$, and $g$. Since we do not know the actual triggering relationship that can be represented by the indicating variables, both \cite{marsan2008extending} and \cite{fox2016spatially} used an EM-type algorithm of \cite{veen2008estimation} for the estimation. In the E step, we calculate the expectation of the complete log-likelihood. Since $\chi_{ij}$ and $\chi_{ii}$ are indicating variables, their expectations are the triggering probabilities of the corresponding pairs of earthquakes, and they can be calculated as
\begin{equation*}
p_{ij} = \frac{\nu(x_i - x_j, y_i - y_j, t_i - t_j; m_j)}{\mu(x_i, y_i) + \sum_{\{j:t_j<t_i\}} \nu(x_i - x_j, y_i - y_j, t_i - t_j; m_j)}
\end{equation*}
if $i > j$, $p_{ij} = 0$ if $i < j$, and
\begin{equation*}
p_{ii} = \frac{\mu(x_i, y_i)}{\mu(x_i, y_i) + \sum_{\{j:t_j<t_i\}} \nu(x_i - x_j, y_i - y_j, t_i - t_j; m_j)}.
\end{equation*}
Now we can make a lower-triangular $N \times N$ triggering probability matrix $P = (p_{ij})_{1 \leq i,j \leq N}$. It is useful in the M step to find the spatially inhomogeneous pattern of the background rate $\mu(x,y)$ and determine the height of each bin in the histogram estimators for $\kappa(m)$, $g_1(\Delta x, \Delta y)$, and $g_2(\Delta t)$. Note that we can give arbitrary numbers as initial values for the triggering probability matrix $P$ \citep{marsan2010new, fox2016spatially}, and iterate the E step and the M step until convergence.

\begin{itemize}
    \item Background rate\\
    In the triggering probability matrix $P$, its diagonal element $p_{ii}$ is a probability that the $i$-th event is a mainshock. Hence, one can estimate the spatially varying background rate in the spatial domain $D$ by a weighted kernel estimator
    \begin{equation}\label{eq:background_kernel}
        \hat{\mu}(x,y)  = \frac{1}{q_{h_1}(x, y|D)\cdot T} \sum_{i=1}^N p_{ii} G_{h_1}(x - x_i, y - y_i),
    \end{equation}
    where $G_{h_1}(\Delta x, \Delta y) = h_1^{-2} \cdot G(\Delta x/h_1, \Delta y/h_1)$ is a (Gaussian) kernel with appropriate bandwidth $h_1$\textcolor{black}{, and $q_{h_1}(x, y|D) = \int\int_D G_{h_1}(x'-x,y'-y)dx'dy'$ is a constant to remedy the edge effect near the boundary of $D$. Our approach to edge correction is detailed in section 1.3 of \cite{diggle2013statistical} and \cite{davies2018tutorial}.}
    
    \item Aftershock productivity\\
    For each event, we can get the expected number of aftershocks through the column-wise summation of $P$ without the diagonal element. So, what we need to do is finding a function $\kappa(m)$ which best explains the relationship between the magnitude $m_j$ and the event-wise productivity $\sum_{i=j+1}^N p_{ij}$. For a given bin $[\omega_1, \omega_2)$ in the magnitude domain, we estimate the height of the histogram by 
    \begin{equation*}
        \hat{\kappa}(m) = \frac{\sum_{\{j: \omega_1 \leq m_j < \omega_2\}} \sum_{i=j+1}^N p_{ij}}{\sum_{\{j: \omega_1 \leq m_j < \omega_2\}} 1}
    \end{equation*}
    when $m$ in on a bin $[\omega_1, \omega_2)$.
    
    \item Triggering density\\
    Spatial triggering density $g_1$ is assumed to be isotropic, which makes it expressed as ($\Delta \mathbf{s} = \sqrt{\Delta x^2 + \Delta y^2}$)
    \begin{equation*}\label{eq:trig_spat_iso}
    g_1(\Delta x, \Delta y) = \frac{g_{01}(\Delta \mathbf{s})}{2 \pi \Delta \mathbf{s}}, 
    \end{equation*}
    by a change-of-variable to the polar coordinate and integrating out the angular variable. Now we can obtain the histogram estimators for $g_{01}$ and $g_2$ in a similar manner. For example, let us assume that we want to find the heights on the bins of the histogram which is estimating $g_2$. Then, we have
    $$\hat{g}_2(\Delta t) = \frac{\sum_{\{(i,j):\tau_1 \leq t_j - t_i < \tau_2\}} p_{ij}}{(\tau_2 - \tau_1) \sum_{j=1}^{N-1} \sum_{i=j+1}^N p_{ij}}$$
    when $\Delta t$ is on a bin $[\tau_1, \tau_2)$.
    
\end{itemize}

As an extension, we consider a nonparametric ETAS model whose aftershock productivity depends both on magnitude and location of the mainshock. \cite{schoenberg2022nonparametric} suggested a nonparametric method that estimates the aftershock productivity for each event by deriving an analytic form and maximizing the likelihood. But, \cite{schoenberg2022nonparametric} smoothed the aftershock productivities only in a magnitude domain without considering their spatial variability. Furthermore, it requires the invertibility of an $(N-1) \times (N-1)$ possibly ill-conditioned lower-triangular matrix G whose $(i,j)$-th element $G_{ij}$ is $g(x_{i+1} - x_j, y_{i+1} - y_j, t_{i+1} - t_j)$ if $i \geq j$, and 0 otherwise. 

Nonparametric ETAS model with anisotropic triggering structure was first suggested by \cite{gordon2021nonparametric}. It estimates the fault direction of each earthquake and assumes that aftershocks occur at varying angles to the estimated direction. As a result, the spatial triggering density is a function of both the relative angle and the spatial lag $\Delta \mathbf{s}$. \cite{gordon2021nonparametric} estimated this bivariate function using a histogram estimator. However, its locally constant form can lead to undesired bumps depending on how partition was done. \textcolor{black}{This problem can be alleviated by kernel methods. \cite{mohler2011self} used the kernel smoothing method to estimate the Hawkes process models. \cite{zhuang2019semiparametric} estimated periodic background rate by introducing so-called relaxation parameters and using kernel-based residual analysis. In this paper, we adopt the kernel smoothing approaches of \cite{mohler2011self} and \cite{fox2016spatially}. We estimate the aftershock productivity and the triggering density as in \eqref{eq:background_kernel}.}

\section{Flexible Hawkes Process Models}\label{sec:3}
This section proposes a new kernel-based nonparametric ETAS model, which has three new attributes for flexibility. It can be expressed by a following conditional intensity function
\begin{equation}\label{eq:proposed_conditional_intensity}
    \lambda(x,y,t|\mathcal{H}_t) = \mu(x,y) + \sum_{\{j:t_j < t\}} \alpha(x_j, y_j) \kappa(m_j) g(x - x_j, y - y_j, t - t_j; \eta, \theta).
\end{equation}
Here, $\alpha(x,y)$ is for the first new attribute. It is a multiplicative correction term which allows the aftershock productivity to change over space. Second, our proposed triggering density $g(\Delta x,\Delta y,\Delta t;\eta,\theta)$ has two new parameters to reflect the anisotropy in the aftershock spatial pattern. Parameters $\eta \geq 1$ and $\theta$ determine the eccentricity and the major axis direction of the elliptic spatial pattern of aftershocks, respectively (see \eqref{anisotropy}). Third, we also assume space-time non-separability for the possible interaction between spatial and temporal lags. 

For the estimation of $\mu$, $\alpha$, $\kappa$, and $g$ in \eqref{eq:proposed_conditional_intensity}, we use the kernel methods to allow the estimates to vary smoothly over space, time, or magnitude. Smooth estimator is more advantageous for the global aftershock productivity $\kappa(m)$ than the other components. According to Gutenberg–Richter law, the frequency of earthquakes decreases exponentially as the magnitude increases \citep{gutenberg1941seismicity}. Therefore, earthquakes with large magnitudes are relatively less frequent compared to those with smaller magnitudes. In histogram based methods, one may account for this by assigning wide bins for large magnitudes. However, the aftershock productivity is expected to increase faster as magnitude gets larger. This implies that the constant productivity may be inappropriate especially on the wide bins with large magnitudes. Hence, we propose to estimate aftershock productivity by kernel smoothing of event-wise productivity:
\begin{equation*}
    \hat{\kappa}(m) = \frac{\sum_{j=1}^{N-1} (\sum_{i=j+1}^{N} p_{ij}) G_{h_2}(m - m_j)}{\sum_{j=1}^{N-1} G_{h_2}(m - m_j)},
\end{equation*}
where $G_{h_2}(\Delta m) = h_2^{-1} G(\Delta m / h_2)$ is a (Gaussian) kernel with appropriate bandwidth $h_2$. 

The rest of this section illustrates how three new attributes are estimated nonparametrically by dividing them into three subsections. A new MISD algorithm that incorporates these new features can be found in the Appendix A.

\subsection{Spatially varying aftershock productivity}

This subsection contains our new work, in which we propose a nonparametric ETAS model which can explain the aftershock productivity with location as well as magnitude. Our approach is in common with \cite{schoenberg2022nonparametric} in the point that $\kappa(m)$ is obtained by smoothing the eventwise productivity, but we do not need the invertibility of matrix $G$. To allow spatially varying features, we introduce a regional aftershock productivity correction factor $\alpha(x,y)$ and multiply it to the global productivity function $\kappa(m)$. To this end, we focus on the discrepancy between the magnitude-based global aftershock productivity $\kappa(m_j)$ and the eventwise aftershock productivity $\sum_{i=j+1}^N p_{ij}$ for $j = 1, 2, \cdots, N-1$.
Since we are using a kernel smoothing of $\sum_{i=j+1}^{N-1} p_{ij}$ in the magnitude domain to get $\kappa(m_j)$, the estimated number of triggered events can be obtained by summing up either of them for the entire events in the catalog. Therefore, the ratio
\begin{equation*}\label{eq:total_aftershock_ratio}
    A^* = \frac{\sum_{j=1}^{N-1} \sum_{i=j+1}^N p_{ij}}{\sum_{j=1}^{N-1} \kappa(m_j)}
\end{equation*}
would have a value close to unity. Note that it is hard to have the equation $A^* = 1$ hold because  $\kappa(m)$ is estimated with a kernel method.

However, on a local spatial neighborhood, this ratio will fluctuate from the constant $A^*$  if there is a tendency that $\kappa(m)$ overestimates (or underestimates) the aftershock productivity compared to the eventwise productivity. Let $\mathcal{J}$ denote such a region (or a set of event indexes occurring on that region) on which the actual aftershock productivity is underestimated by $\kappa(m)$. Then the ratio $A_{\mathcal{J}}^* = \sum_{j\in \mathcal{J}} \sum_{i=j+1}^N p_{ij} / \sum_{j\in \mathcal{J}} \kappa(m_j)$ which is restricted on the region $\mathcal{J}$ becomes substantially larger than $A^*$. So, if $A_{\mathcal{J}} = A^*_{\mathcal{J}} / A^* > 1$, it would mean that the earthquakes occurring on the region $\mathcal{J}$ have higher aftershock productivity compared to the rest part on average. If $A_{\mathcal{J}} < 1$, it would mean the opposite. Furthermore, if we define $A_{D \setminus \mathcal{J}}$ similarly as we did for $A_{\mathcal{J}}$ on the region other than $\mathcal{J}$ in the spatial domain $D$, we have
\begin{equation*}
    \frac{\sum_{j =1}^{N-1} \sum_{i=j+1}^N p_{ij}}{\sum_{j\in \mathcal{J}} A_{\mathcal{J}} \kappa(m_j) + \sum_{j\in D \setminus \mathcal{J}} A_{D \setminus \mathcal{J}} \kappa(m_j)} = A^*,
\end{equation*}
and this implies that we can understand $A_{\mathcal{J}}$ as a regional productivity correction factor which reflects the geological characteristics implicitly on the region $\mathcal{J}$.

Now there remains the problem of distinguishing the region $\mathcal{J}$ from the other parts. In reality, however, the aftershock productivity can vary gradually over space as a result of many and sometimes unknown  factors such as different tectonic, geological, and stress states.  Fortunately, we can bypass this problem of uncovering the underlying structure by not partitioning the space one from the other but instead calculating the productivity correction factor on each point $(x,y)$.

To this end, we consider local averages of the eventwise aftershock productivity $\sum_{i=j+1}^N p_{ij}$ and the global aftershock productivity $\kappa(m_j)$ based on the same (Gaussian) kernel $G_{h_3}(\Delta x, \Delta y) = h_3^{-2} \cdot G_{h_3}(\Delta x / h_3, \Delta y / h_3)$ with appropriate bandwidth $h_3$. Then we can calculate a spatially varying ratio as a function of longitude $x$ and latitude $y$,
\begin{equation}\label{eq:alpha_star_est}
    \alpha^*(x,y) = \frac{\sum_{j =1}^{N-1} (\sum_{i=j+1}^N p_{ij}) G_{h_3}(x - x_j, y - y_j)}{\sum_{j =1}^{N-1} \kappa(m_j) G_{h_3}(x - x_j, y - y_j)}.
\end{equation}
The aftershock productivity correction factor can then be obtained on each point by
\begin{equation}\label{eq:alpha_est}
    \alpha(x,y) = \alpha^*(x,y) / A^*,
\end{equation}
and this is multiplied to the value of $\kappa$ of an event at the corresponding location.

Estimation of $\alpha(x,y)$ can be incorporated in the iterative nonparametric method in Section \ref{sec:2}.  Once the estimate of ${\kappa}(m)$ is obtained, we can estimate $\alpha(x,y)$ using  Equations \eqref{eq:alpha_star_est} and \eqref{eq:alpha_est}. After that, we can update the triggering probability matrix $P$ by
\begin{equation*}
{p}_{ij} = \frac{{\alpha}(x_j, y_j){\kappa}(m_j){g}(x_i - x_j, y_i - y_j, t_i - t_j)}{{\mu}(x_i, y_i) + \sum_{\{j:t_j<t_i\}} {\alpha}(x_j, y_j){\kappa}(m_j){g}(x_i - x_j, y_i - y_j, t_i - t_j)}
\end{equation*}
for $i > j$, and
\begin{equation*}
{p}_{ii} = \frac{{\mu}(x_i, y_i)}{{\mu}(x_i, y_i) + \sum_{\{j:t_j<t_i\}} {\alpha}(x_j, y_j){\kappa}(m_j){g}(x_i - x_j, y_i - y_j, t_i - t_j)}.
\end{equation*}
In this way, we obtain new estimates of $\kappa(m)$ and $\alpha(x,y)$ based on these probabilities, and the algorithm iterates until the convergence of the triggering probability matrix.

\subsection{Anisotropic spatial triggering mechanism}\label{subsec:3.2}

We also propose a nonparametric ETAS model whose triggering density can account for the elliptic feature similarly as in \cite{ogata1998space}. We assume that the aftershocks are equally likely to occur if their Mahalanobis distances $ \sqrt{(\Delta x\ \Delta y) S_{\eta\theta}^{-1} (\Delta x\ \Delta y)^T}$ are the same for a matrix
\begin{equation}
    S_{\eta\theta} = \begin{pmatrix}
    \cos\theta & -\sin\theta \\
    \sin\theta & \cos\theta
    \end{pmatrix}
    \begin{pmatrix} \eta & 1 \\
    1 & 1/\eta
    \end{pmatrix}
    \begin{pmatrix}
    \cos\theta & \sin\theta \\
    -\sin\theta & \cos\theta
    \end{pmatrix}.\label{anisotropy}
\end{equation}
The parameter $\eta \geq 1$ represents the ratio between the major and minor axes of the ellipse, and $\theta$ denotes the angle between the major axis and a virtual horizontal line. So, the triggering density can reflect the anisotropy by adding the parameters $\eta$ and $\theta$ as
$$g(\Delta x, \Delta y, \Delta t;\eta, \theta) = g_{1}(\Delta x, \Delta y;\eta, \theta) g_2(\Delta t),$$
and measuring the spatial lags with the Mahalanobis distance.

\begin{figure}[h!]
     \centering
     \begin{subfigure}[b]{0.3\textwidth}
         \centering
         \includegraphics[width=\textwidth]{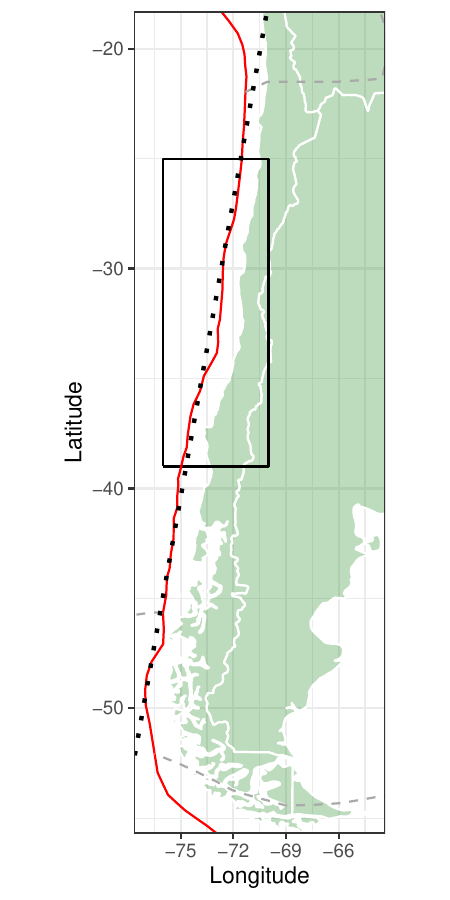}
         \caption{}
         \label{fig:chile_approxline}
     \end{subfigure}
     \hspace{2em}
     \begin{subfigure}[b]{0.3\textwidth}
         \centering
         \includegraphics[width=\textwidth]{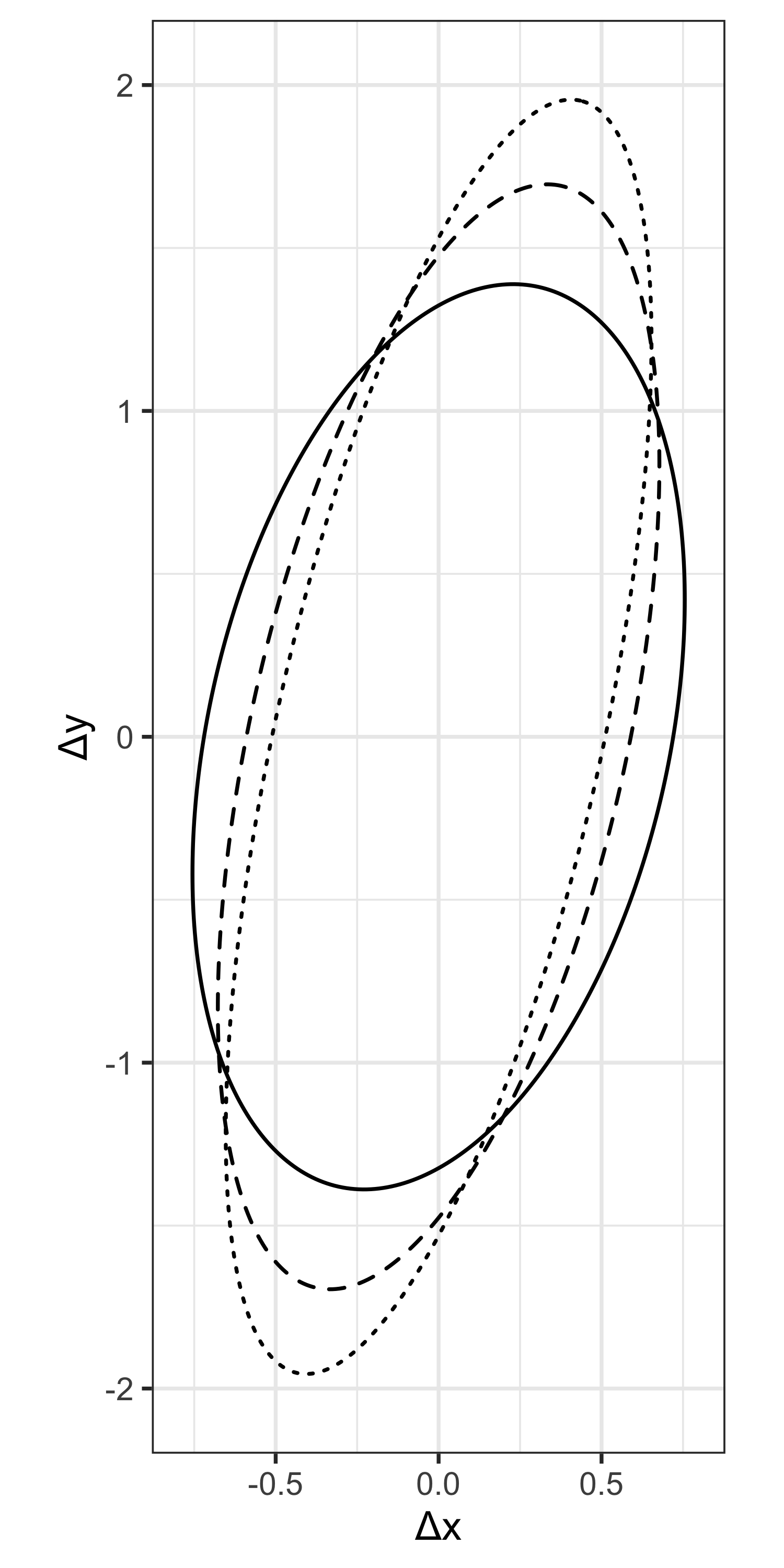}
         \caption{}
         \label{fig:chile_elip}
     \end{subfigure}
     \caption{(a) Plate boundary is approximated by a dotted black line. The box represents the spatial domain of interest. The solid red line is the subducting portion of the plate boundary, and the dashed gray lines are non-subducting boundaries. (b) Different values of $\eta = 2, \, 3, \, 4$ are represented by solid, dashed, and dotted ellipses, respectively.}
     \label{fig:chile_anisotropy}
\end{figure}

As in many other cases of point process data, it may be challenging to find out the underlying geometry of the aftershock triggering mechanism. Even if the ETAS model fits the data very well, we get $O(N^2)$ pairs of probabilistic relationships. As a result, this paper makes use of the fact that most of the earthquakes are caused by the relative motion of planar fault surfaces \citep{lay1995modern,Lietal2018}. Since dominant fault strikes tend to follow the direction of the nearby major plate boundary, we can approximate the fault strikes with a straight line if we confine the spatial domain small enough. Figure \ref{fig:chile_approxline} illustrates the approximated subducting boundary in the Chile region with dotted line. The black box depicts the region of our interest, the red line is the portion of the subducting plate boundary, and the dashed gray lines are non-subducting boundaries. The slope of the dotted line can be easily calculated because boundary information is given in a piecewise linear form. We perform a linear regression with a midpoint on each segment with corresponding segment length as a weight. As a result, we get $\theta = 75.64^\circ$ as a slope angle with respect to the horizontal direction (or the East direction). On the other hand, Figure \ref{fig:chile_elip} shows three ellipses of $$(\Delta x\ \Delta y) S_{\eta\theta}^{-1} (\Delta x\ \Delta y)^T = 1.$$ These ellipses share the same direction $\theta=75.64^\circ$, but have different axial ratios $\eta = 2, \, 3, \, 4$  represented by solid, dashed, and dotted curves, respectively. This implies that a larger value of $\eta$ is required if the aftershocks are more likely to concentrate along the line of direction $\theta$. However, the degree of the anisotropy $\eta$ is difficult to determine directly. Therefore, we suggest fitting the ETAS model with a range of $\eta$ values and then choosing the one that produces the most accurate forecasts. 

\subsection{Space-time interaction in aftershocks} 
A common assumption on the triggering density $g(\Delta x, \Delta y, \Delta t)$ is that it can be decomposed separately into spatial and temporal components as in \eqref{eq:trig_separable}. However, space-time separability can reflect the space-time interaction in the aftershock occurrences. In other words, it makes the triggering effect of a mainshock have the same temporal decay rate at two locations with different spatial lags. For this reason, we propose to use space-time non-separability.

The triggering density can be expressed as a bivariate function $$g_0(\Delta \mathbf{s}, \Delta t) =  2\pi \Delta \mathbf{s}~g(\Delta x, \Delta y, \Delta t).$$
For fast computation, a binned kernel estimator \citep{silverman1982algorithm, wand1994fast} is used by dividing the domain of $g_0$ into an equally-spaced grid. However, we are more interested in the region of small $\Delta \mathbf{s}$ and $\Delta t$ because aftershocks are more likely to occur when they are close to the location and time of the triggering mainshock. Before using the binned kernel estimator, we log-transform and standardize these spatial and temporal lags as 
\begin{equation*}\label{eq:lag_log_std}
        \Delta \mathbf{s}_{ij}^* = \log(\Delta \mathbf{s}_{ij} + 1) / \sigma_s, \, \Delta t_{ij}^* = \log(\Delta t_{ij} + 1) / \sigma_t,
    \end{equation*}
where $\Delta \mathbf{s}_{ij}$ and $\Delta t_{ij}$ are the spatial and the temporal lags of the $i$-th and the $j$-th events when the latter precedes the former ($1 \leq j < i \leq N$), and $\sigma_s$ and $\sigma_t$ are standard deviations of $\log(\Delta \mathbf{s}_{ij} + 1)$ and $\log(\Delta t_{ij} + 1)$, respectively. For each of these lags, there is a weight $p_{ij}$ which tells whether the lag is for a pair of events that are actually in a mainshock-aftershock relationship. As a result, we use a weighted kernel density estimator
\begin{equation*}
    \hat{g}^*_0(\Delta \mathbf{s}^*,\Delta t^*) = \frac{\sum_{i>j} p_{ij} G_{h_4} (\Delta \mathbf{s}^* - \Delta \mathbf{s}^*_{ij}, \Delta t^* - \Delta t^*_{ij})}{q_{h_4}(\Delta \mathbf{s}^*, \Delta t^*|\mathbb{R}^{+} \times \mathbb{R}^{+})\sum_{i>j} p_{ij} \sum_{i>j}1},
\end{equation*}
where $G_{h_4}(\cdot, \cdot) = h_{4}^{-2} \cdot G(\cdot / h_4, \cdot / h_4)$ is a bivariate (Gaussian) kernel with appropriate bandwidth $h_4$\textcolor{black}{, and $q_{h_4}(\Delta \mathbf{s}^*, \Delta t^*|\mathbb{R}^{+} \times \mathbb{R}^{+}) = \int\int_{\mathbb{R}^{+} \times \mathbb{R}^{+}} G_{h_4}(\Delta \mathbf{s}'-\Delta \mathbf{s}^*,\Delta t'-\Delta t^*)d\Delta \mathbf{s}'d\Delta t'$ is a constant to remedy the edge effect near the boundary as in \eqref{eq:background_kernel}.} By change-of-variable, we can revert this back to original unit as
\begin{equation*}
 \hat{g}_0(\Delta \mathbf{s},\Delta t) = \frac{\hat{g}^*_0(\Delta \mathbf{s}^*,\Delta t^*)}{\sigma_s \sigma_t \exp(\sigma_s \Delta \mathbf{s}^* + \sigma_t \Delta {t}^*)}.
\end{equation*}

\section{Application to Earthquake Data}\label{sec:4}

We now apply our newly proposed approaches to multiple earthquake catalogs (with major earthquake activities). Catalogs from five time periods in two different regions are investigated, and several variants of kernel-based ETAS models are evaluated. Fitted results from the best model for each case are then compared to those from the ETAS model, which does not assume spatially varying productivity, anisotropy, and space-time interaction in aftershock occurrences. Finally, we compare how the estimated background rate changes before and after major earthquakes.

\subsection{Data specification}

\begin{figure}[h!]
     \centering
     \begin{subfigure}[b]{0.3\textwidth}
         \centering
         \includegraphics[width=\textwidth]{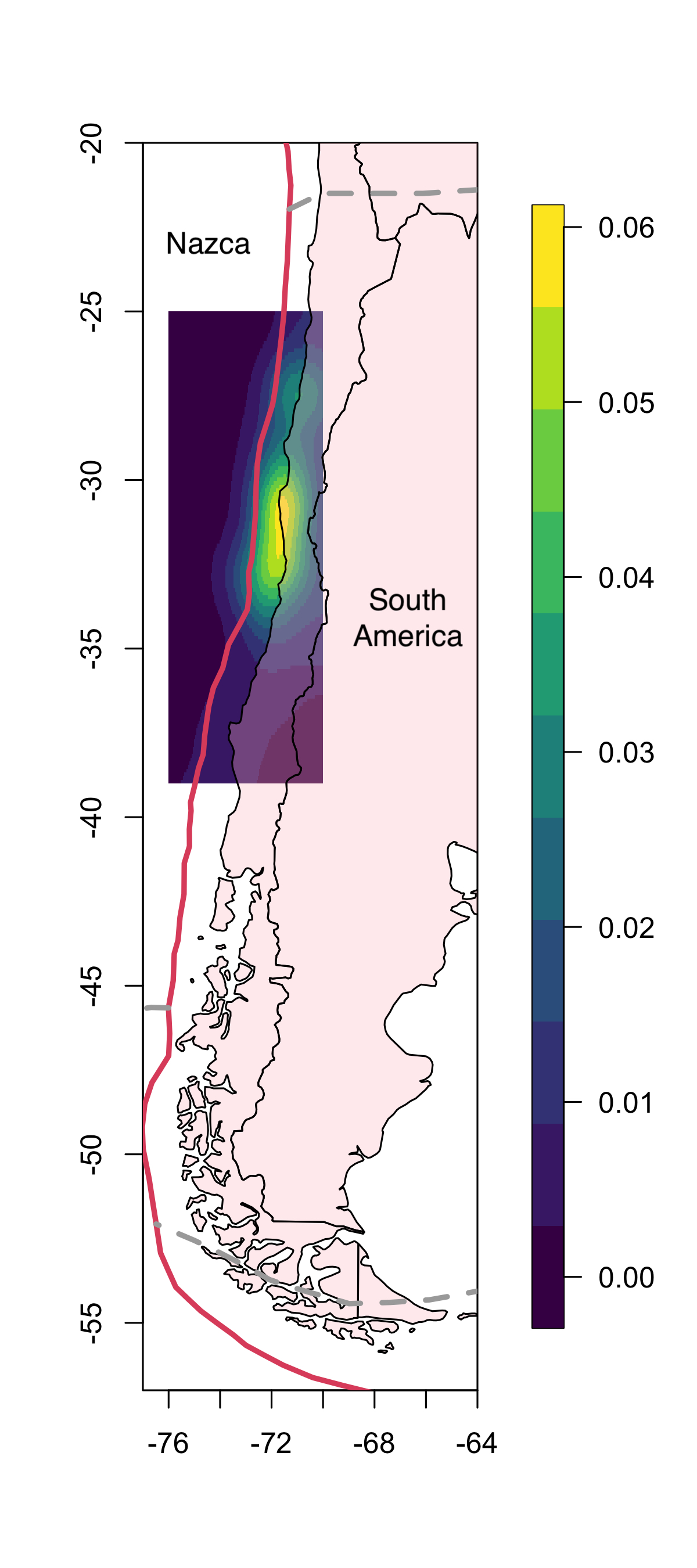}
         \caption{}
         \label{fig:sdomain_chile_a}
     \end{subfigure}
     \hfill
     \begin{subfigure}[b]{0.3\textwidth}
         \centering
         \includegraphics[width=\textwidth]{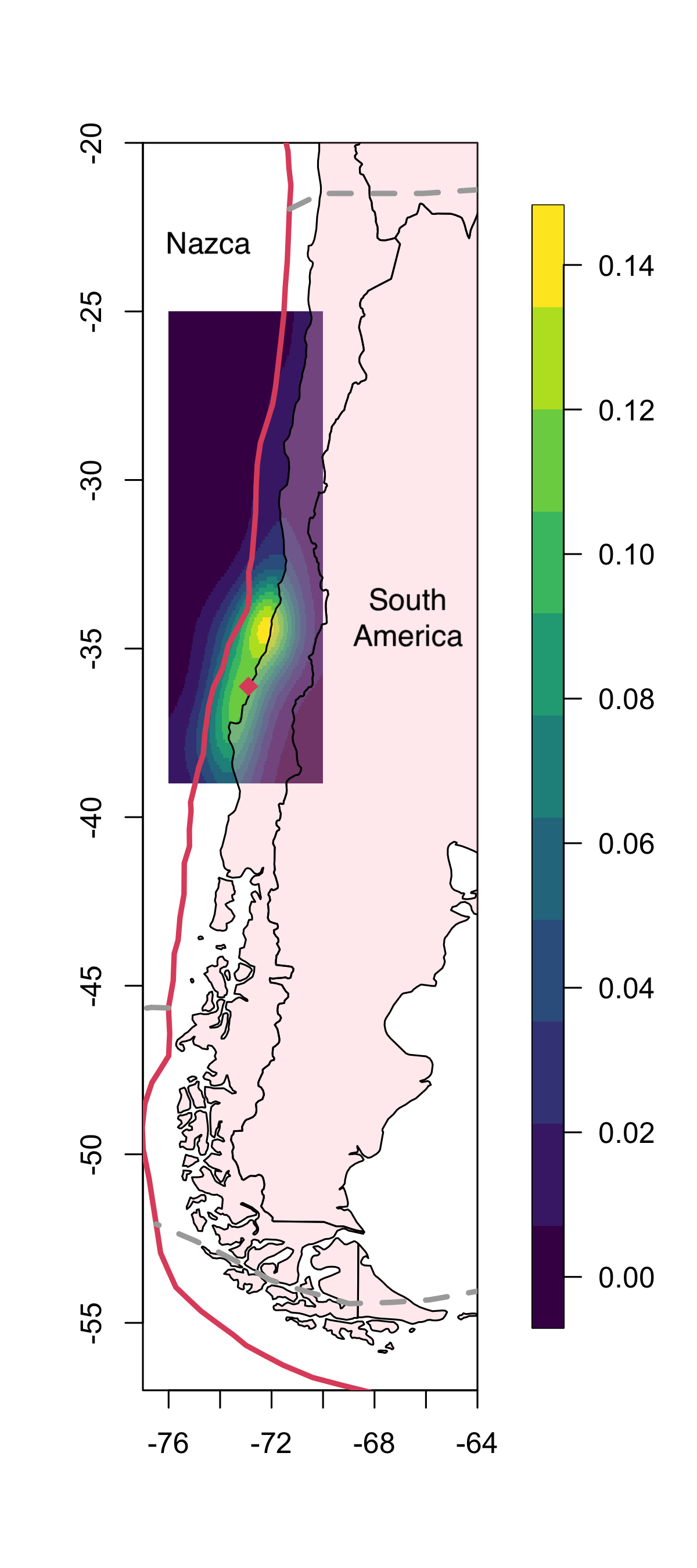}
         \caption{}
         \label{fig:sdomain_chile_b}
     \end{subfigure}
     \hfill
     \begin{subfigure}[b]{0.3\textwidth}
         \centering
         \includegraphics[width=\textwidth]{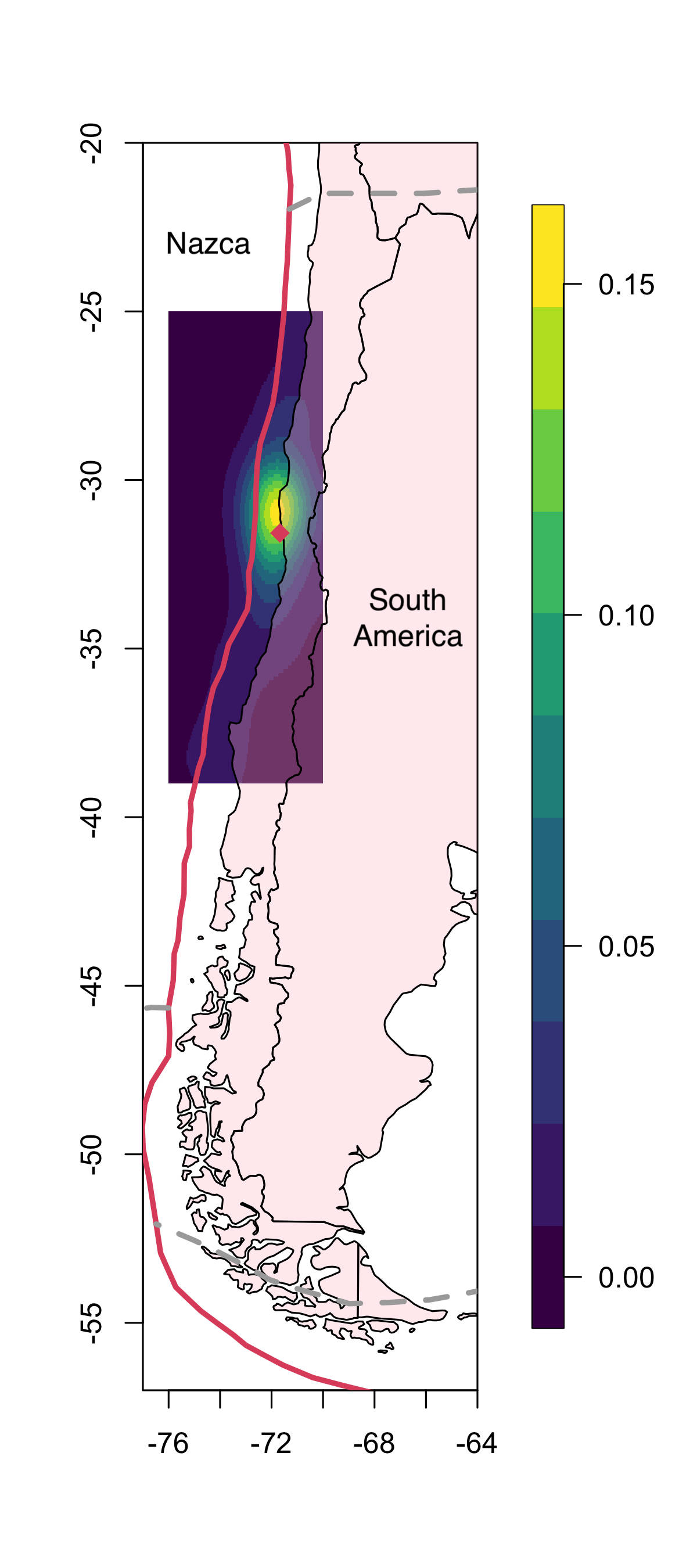}
         \caption{}
         \label{fig:sdomain_chile_c}
     \end{subfigure}
     \caption{Scaled spatial intensities $T^{-1}\cdot \lambda(x,y)$ of the catalogs (a) Chile A, (b) Chile B, and (c) Chile C. The solid red line represents the portion of the subducting plate boundary, and the dashed gray line represents the non-subducting boundary. The diamond symbol ($\MyDiamond$) marks the epicenter of the major earthquake.}
     \label{fig:stdomain_chile}
\end{figure}

\begin{figure}[h!]
     \centering
     \begin{subfigure}[b]{0.45\textwidth}
         \centering
         \includegraphics[width=\textwidth]{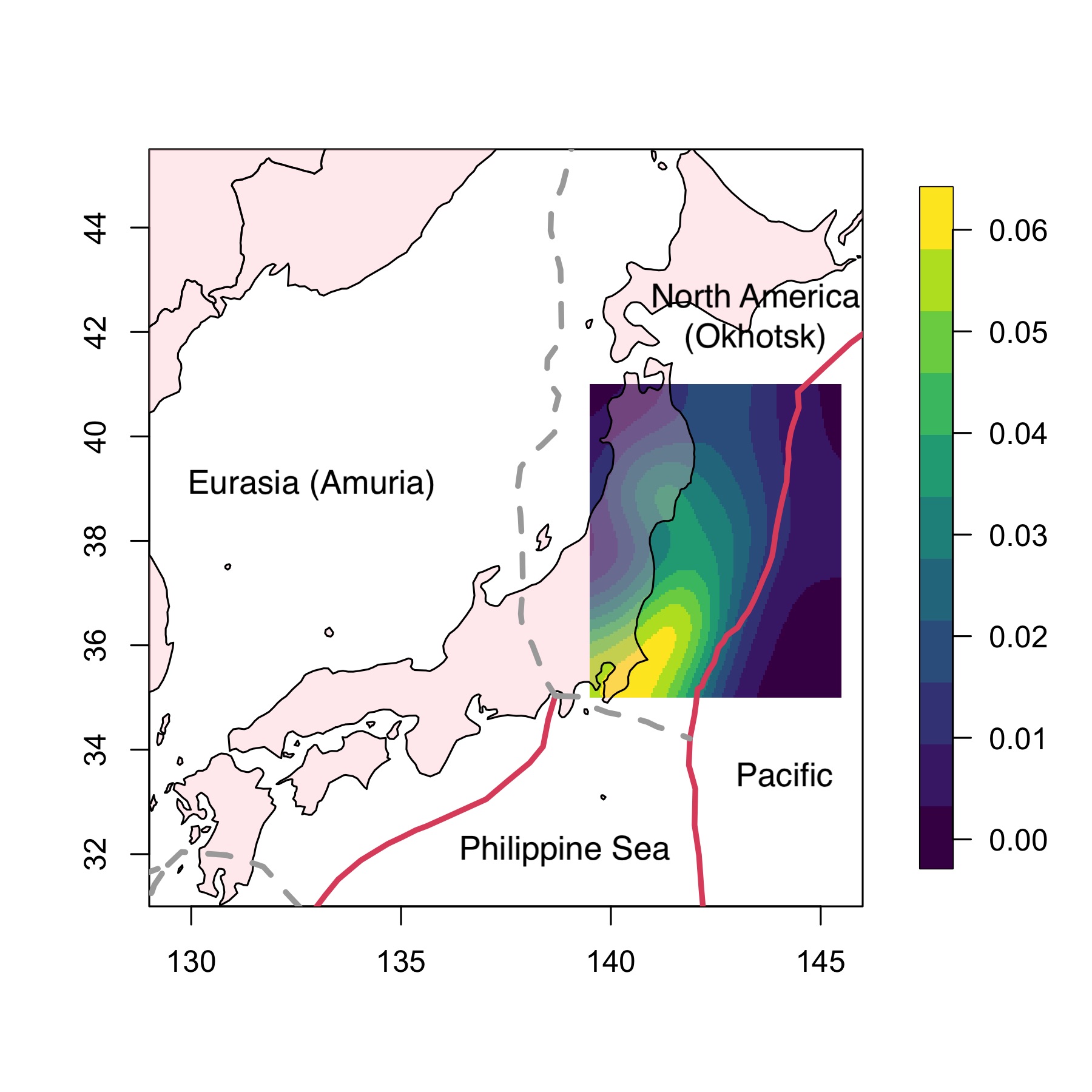}
         \caption{}
         \label{fig:sdomain_japan_a}
     \end{subfigure}
     \hfill
     \begin{subfigure}[b]{0.45\textwidth}
         \centering
         \includegraphics[width=\textwidth]{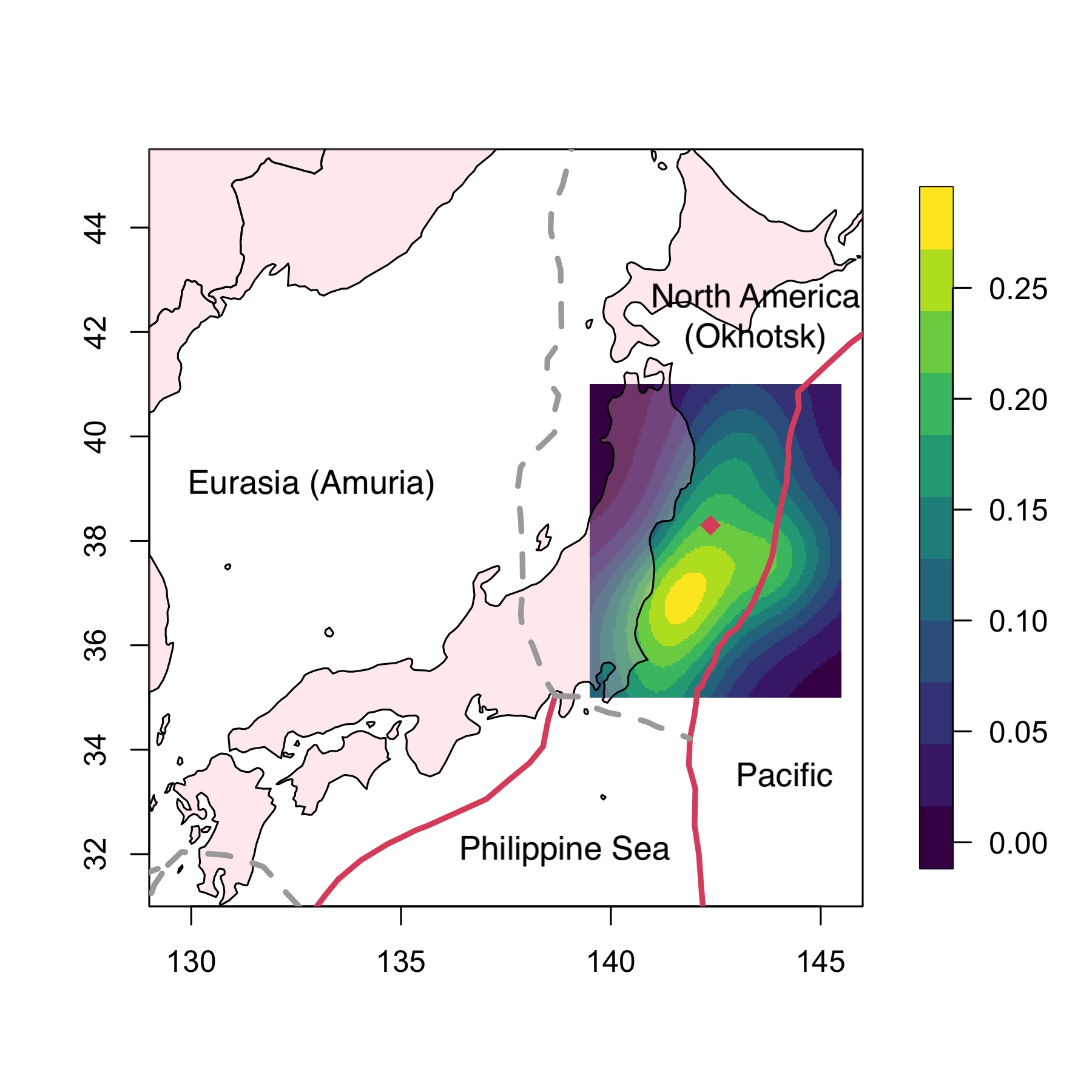}
         \caption{}
         \label{fig:sdomain_japan_b}
     \end{subfigure}
     \caption{Scaled space-time intensities $T^{-1}\cdot \lambda(x,y)$ of the catalogs (a) Japan A and (b) Japan B. The solid red line represents the subducting plate boundary, and dashed gray line represents the non-subducting boundary. Diamond symbol ($\MyDiamond$) marks the epicenter of the major earthquake. See \cite{zheng2006low} for the discussion on the minor tectonic plates in this region.}
     \label{fig:stdomain_japan}
\end{figure}

We examine the proposed approaches on earthquake data from Chile and Japan regions. Tectonic plates subduct under the ocean near these countries to drive the seismic activities, \textcolor{black}{and we determine the spatial domains so that the majority of the earthquakes are located away from the boundary in order to reduce the problem of edge effect.} The spatial domain near Chile is selected as $\{(L,l): L \in [-39^\circ, -25^\circ] ,\, l \in [-76^\circ, -70^\circ]\}$ (Figure \ref{fig:stdomain_chile}), where $L$ and $l$ denote the latitude and longitude, respectively. In this area, the Nazca plate in the Pacific Ocean subducts eastward under South America. The spatial domain near Japan is $\{(L,l): L \in [35^\circ, 41^\circ], \, l \in [139.5^\circ, 145.5^\circ]\}$ (Figure \ref{fig:stdomain_japan}), where the western part of the Pacific plate subducts under Japan.

For the temporal domains, we choose three observation periods for the Chile region and two for the Japan region before and after the recent large earthquakes. An earthquake of magnitude 8.8 occurred near Chile on February 27, 2010, and another of magnitude 8.3 occurred on September 16, 2015. In the Japan region, an earthquake of magnitude 9.1 occurred on March 11, 2011. Table \ref{tab:catalog_summary} summarizes the earthquake data catalogs analyzed in this paper\textcolor{black}{, which excludes the deep earthquakes whose focal depths are over 100km and cuts off the small earthquakes with magnitudes less than 4.0}. Each of the five catalogs lasts approximately six years, with the last year of each as a forecast period for evaluating the  flexible ETAS models. Catalogs from the Chile region are labeled as `Chile A,' `Chile B,' and `Chile C' in chronological order, and similarly for the Japan region as `Japan A' and `Japan B.' Figures \ref{fig:stdomain_chile} and \ref{fig:stdomain_japan} illustrate the scaled spatial intensities $T^{-1} \cdot \lambda(x,y)$ for all earthquakes (which do not distinguish the mainshocks and the aftershocks) in the catalogs from Chile and Japan, respectively. Note that the length of training period, $T$, is divided to get comparable values for different catalogs.

\begin{table}[h!]
\centering
\caption{Summary of the earthquake catalogs from the Chile region $\{(L,l): L \in [-39^\circ, -25^\circ] ,\, l \in [-76^\circ, -70^\circ]\}$ and the Japan region $\{(L,l): L \in [35^\circ, 41^\circ], \, l \in [139.5^\circ, 145.5^\circ]\}$ ($L$: latitude, $l$: longitude)}
\label{tab:catalog_summary}
\resizebox{\textwidth}{!}{%
\begin{tabular}{@{}ccc@{}}
\toprule
Catalog & Training period      & Forecast period      \\
\midrule
Chile A      & 01/01/2001 - 12/31/2005 (1273 events) & 01/01/2006 - 12/31/2006 (296 events) \\
Chile B      & 02/27/2010 - 09/15/2014 (2882 events) & 09/16/2014 - 09/15/2015 (228 events) \\
Chile C      & 09/16/2015 - 09/15/2020 (2291 events) & 09/16/2020 - 09/15/2021 (261 events) \\
Japan A      & 01/01/2003 - 12/31/2007 (875 events) & 01/01/2008 - 12/31/2008 (452 events)\\
Japan B      & 03/11/2011 - 03/10/2016 (7001 events) & 03/11/2016 - 03/10/2017 (419 events) \\\bottomrule
\end{tabular}
}
\end{table}

\subsection{Model estimation}\label{subsec:4.2}
We use prefixes to name the models considered. For the aftershock productivity, V stands for spatially varying $\alpha(x,y)$ and C for constant (i.e. $\alpha(x,y)=1$). For the triggering function, N stands for space-time non-separable $g$, and S for separable $g$. Regarding the degree of anisotropy, we compare the values $\eta = 1,\, 2,\, 3,\, 4,\, \cdots$ for each catalog. These are appended after the prefixes  as VN-1:1, VN-2:1, VN-3:1, and so forth. For the direction of the anisotropy pattern, we orient the major axis to approximate the plate boundary in each region. We use the coordinates of the boundary and apply a weighted linear regression as described in subsection \ref{subsec:3.2} to determine the local plate boundary orientation. The angle of the estimated regression line can be expressed as a counterclockwise angle from the horizontal line (the east direction). We have obtained $\theta = 75.64^\circ$ for three catalogs from Chile and $\theta = 65.78^\circ$ for two catalogs from Japan. But, we have to note that the degree of anisotropy $\eta$ can be different for two catalogs with same spatial domain because the region of active seismic activity may vary over time from period to period, shown in Figure \ref{fig:stdomain_chile}. 

For the estimation of $\mu(x,y)$ and $\alpha(x,y)$, a kernel method with fixed bandwidth has its limitation due to the clustering structure of epicenters. A small bandwidth results in a noisy estimate of the region with few earthquakes, while a large bandwidth blurs out the patterns in the seismically active region. To alleviate this problem, we adjust the kernel bandwidth by adopting the square root rule of \cite{abramson1982bandwidth}, which is used for the intensity function estimation due to its small bias \citep{davies2018fast, gonzalez2022adaptive}. We first estimate the weighted kernel density $f_0$ with a Gaussian kernel with bandwidth $h_0 = 0.5^\circ$. Then we adjust the bandwidth on each epicenter $(x_i, y_i)$ to be $h_i = h_0 f_0(x_i, y_i)^{-1/2} \gamma^{-1}$, where $\gamma$ is the geometric mean of $f(x_i, y_i)^{-1/2}$. This allows the kernels centered on the region of sparse earthquakes to have wider bandwidths, while the kernels centered on the densely observed region to have narrower bandwidths. Since the aftershock productivity function $\kappa(m)$ is neither density nor intensity, we use the $k$-th nearest epicenter to select the kernel bandwidth. Here, we determine the value of $k$ by using the leave-one-out cross validation with a least-squares criterion. However, the triggering density $g_0^*(\Delta \mathbf{s}^*, \Delta t^*)$ is estimated with a Gaussian kernel with a fixed bandwidth $0.2^\circ$ to avoid the computational burden resulting from $O(N^2)$ pairs of spatial and temporal lags when there are $N$ earthquakes.

\subsection{Model comparison}\label{subsec:4.3}
The results of all the models considered are compared based on their daily forecast accuracy. We first fit the  ETAS models to the historical events that occurred during the training period of each catalog. We then evaluate the conditional intensity \eqref{eq:proposed_conditional_intensity} on the midpoints of $0.1^\circ \times 0.1^\circ$ cells over the spatial domain at the beginning of every day during the forecast period. Now we produce forecast based on a threshold value for the conditional intensity. If the conditional intensity exceeds a certain threshold, we forecast that one or more earthquakes might occur in the cell within 24 hours following the midnight. Otherwise, we forecast that no earthquakes would occur in the cell in that day. Hence, more space-time cells are forecasted to have one or more earthquakes if the threshold is low and the opposite case is true if the threshold is high. To remediate the effect of an arbitrary threshold, we measure the forecast accuracy based on the area under the curve (AUC) of the receiver operating characteristic (ROC) curve. To be more precise, we calculate the partial AUC by limiting the region of interest for specificity (true negative rate) in the ROC space. This region may vary depending on the circumstances or some expert advice, but we limit the specificity to 50-100\% because sensitivity (true positive rate) reaches nearly 100\% as specificity drops to 50\%. This allows for a better comparison of the models by excluding cases of too low thresholds, which typically result in the increase of false positive forecasts.

\begin{table}[h!]
\centering
\caption{Forecast accuracy of each model measured with partial AUC for the catalogs in the Chile region. The partial AUC with the highest value is bold-faced for each catalog.}
\label{tab:pauc_chile}
\resizebox{\textwidth}{!}{%
\begin{tabular}{@{}c|cccc|cccc|cccccc@{}}
\toprule
            & \multicolumn{4}{c|}{\textbf{Chile A}}                     & \multicolumn{4}{c|}{\textbf{Chile B}}                     & \multicolumn{6}{c}{\textbf{Chile C}}                                                                                  \\ \midrule
            & \textbf{1:1} & \textbf{2:1} & \textbf{3:1} & \textbf{4:1} & \textbf{1:1} & \textbf{2:1} & \textbf{3:1} & \textbf{4:1} & \textbf{1:1} & \textbf{2:1} & \textbf{3:1} & \multicolumn{1}{c}{\textbf{4:1}} & \multicolumn{1}{c}{\textbf{5:1}} &
            \multicolumn{1}{c}{\textbf{6:1}}\\
\textbf{VN} & 0.4129       & {\bf 0.4155}       & 0.4153       & 0.4146       & {\bf 0.3772}       & 0.3744       & 0.3732       & 0.3721       & 0.3840       & 0.3853          & 0.3857       & 0.3860                           & 0.3860         & 0.3853                  \\
\textbf{VS} & 0.4126       & 0.4149       & {0.4146}       & 0.4137       & 0.3759       & 0.3734       & 0.3717       & 0.3706       & 0.3828       & 0.3847          & 0.3845       & {\bf 0.3863}                           & 0.3857         & 0.3848                  \\
\textbf{CN} & 0.4096       & 0.4132       & 0.4137       & 0.4136       & 0.3687       & 0.3682       & 0.3673       & 0.3665       & 0.3791       & 0.3811          & 0.3826       & 0.3828                           & 0.3829        & 0.3827                   \\
\textbf{CS} & 0.4093       & 0.4124       & 0.4130       & 0.4129       & 0.3691       & 0.3681       & 0.3670       & 0.3657       & 0.3785       & 0.3805          & 0.3819       & 0.3822                           & 0.3825       & 0.3823                    \\ \bottomrule
\end{tabular}%
}
\end{table}

\begin{figure}[hbt!]
     \centering
     \begin{subfigure}[b]{0.3\textwidth}
         \centering
         \includegraphics[width=\textwidth]{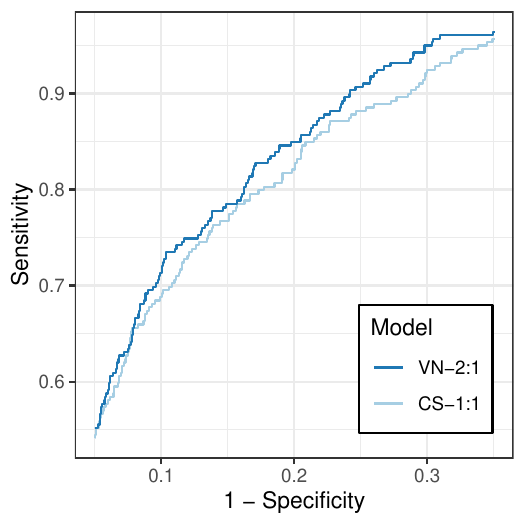}
         \caption{Chile A}
         \label{fig:proc_chile_a}
     \end{subfigure}
     \hfill
     \begin{subfigure}[b]{0.3\textwidth}
         \centering
         \includegraphics[width=\textwidth]{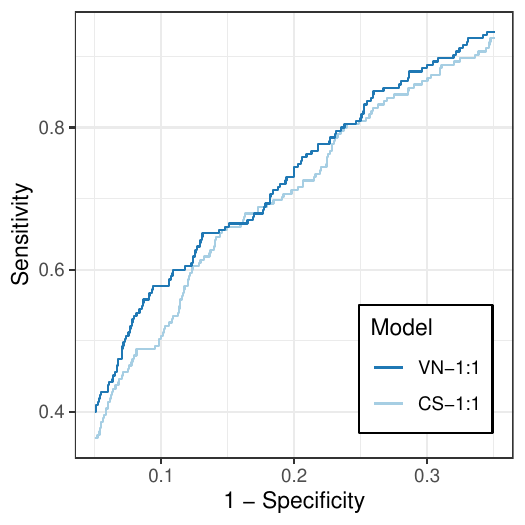}
         \caption{Chile B}
         \label{fig:proc_chile_b}
     \end{subfigure}
     \hfill
     \begin{subfigure}[b]{0.3\textwidth}
         \centering
         \includegraphics[width=\textwidth]{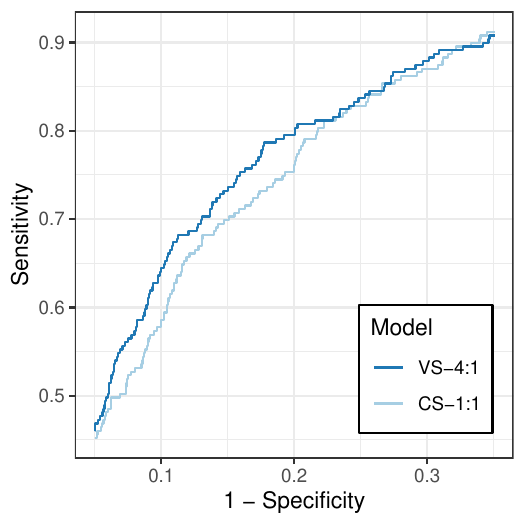}
         \caption{Chile C}
         \label{fig:proc_chile_c}
     \end{subfigure}
     \caption{Forecast accuracy represented by ROC curves.}
     \label{fig:proc_chile}
\end{figure}

Tables \ref{tab:pauc_chile} and \ref{tab:pauc_japan} summarize the forecast results of the ETAS models from Chile and Japan, respectively. The highest partial AUC from each catalog is bold-faced, and it may be contrasted with a value from CS-1:1 to determine how much the forecast improvement can be achieved by incorporating spatially varying productivity, anisotropy, and space-time interaction in aftershock occurrences. For the catalogs from Chile, models with spatially varying productivity have higher forecast accuracy. The highest partial AUCs are obtained by the models VN-2:1, VN-1:1, and VS-4:1 for the catalogs Chile A, B, and C, respectively.  The improvement is highlighted by the partial ROC curves in Figure \ref{fig:proc_chile}. For the catalog Chile A, VN-2:1 model makes nearly $5$ percent points less false negative forecast compared to CS-1:1 to achieve the sensitivity of $90\%$. For the catalog Chile B, VN-1:1 model improves the sensitivity by nearly $10$ percent points compared to CS-1:1 when the specificity is around $90\%$.
On the other hand, partial AUCs from the catalogs of Japan show relatively little improvement compared to the model CS-1:1. However, we note that small differences in partial AUCs can be actually significant due to the correlation of the ROC curves since we are using the same space-time grid for each catalog. \cite{robin2011proc} addressed this problem and modified the work of \cite{hanley1983method} to implement a bootstrap-based significance test. The test statistic has the form,
$Z = (A_1 - A_2)/sd(A_1 - A_2)$, where $A_1$ and $A_2$ are (partial) AUCs, and it approximately follows a standard normal distribution. For the calculation, we obtain $sd(A_1 - A_2)$ by a stratified bootstrapping of the conditional intensities over the space-time grid. We generate 2000 bootstrap samples with the same size as in the original one and calculate the AUCs for each case.  One-sided tests against the model CS-1:1 give the p-values $1.45\times 10^{-3}$, $5.77\times 10^{-4}$, $2.31\times 10^{-4}$, $4.41\times 10^{-1}$, and $5.62\times 10^{-2}$ in the order of the catalogs from Chile A, B, C and Japan A \& B, respectively. This suggests that there is substantial evidence that flexible models forecast significantly better for the Chile region. It is also notable for the catalog Japan B that the p-value is quite small considering small absolute difference between VN-1:1 and CS-1:1. This is resulting from small variability in the difference between two partial AUCs, which suggests that the proposed method performs better than the existing one in the majority of the cells of the space-time grid.

\begin{table}[hbt!]
\centering
\caption{Forecast accuracy of each model measured with partial AUC for the catalogs from Japan region. Partial AUC with the highest value is bold-faced for each catalog.}
\label{tab:pauc_japan}
\resizebox{11cm}{!}{%
\begin{tabular}{@{}c|cccc|cccc@{}}
\toprule
            & \multicolumn{4}{c|}{\textbf{Japan A}}                     & \multicolumn{4}{c}{\textbf{Japan B}}                      \\ \midrule
            & \textbf{1:1} & \textbf{2:1} & \textbf{3:1} & \textbf{4:1} & \textbf{1:1} & \textbf{2:1} & \textbf{3:1} & \textbf{4:1} \\
\textbf{VN} & 0.3898       & 0.3885       & 0.3864       & 0.3839       & {\bf 0.3679}       & 0.3671       & 0.3665       & 0.3663       \\
\textbf{VS} & {\bf 0.3909}       & 0.3890       & 0.3868       & 0.3846       & 0.3678       & 0.3670       & 0.3664       & 0.3662       \\
\textbf{CN} & 0.3896       & 0.3886       & 0.3867       & 0.3842       & 0.3673       & {0.3668}       & 0.3661       & 0.3655       \\
\textbf{CS} & 0.3908       & 0.3892       & 0.3870       & 0.3847       & 0.3670       & 0.3663       & 0.3656       & 0.3651       \\ \bottomrule
\end{tabular}%
}
\end{table}

\subsection{Result analysis}
\begin{figure}[H]
     \centering
     \begin{subfigure}[b]{0.3\textwidth}
         \centering
         \includegraphics[width=\textwidth]{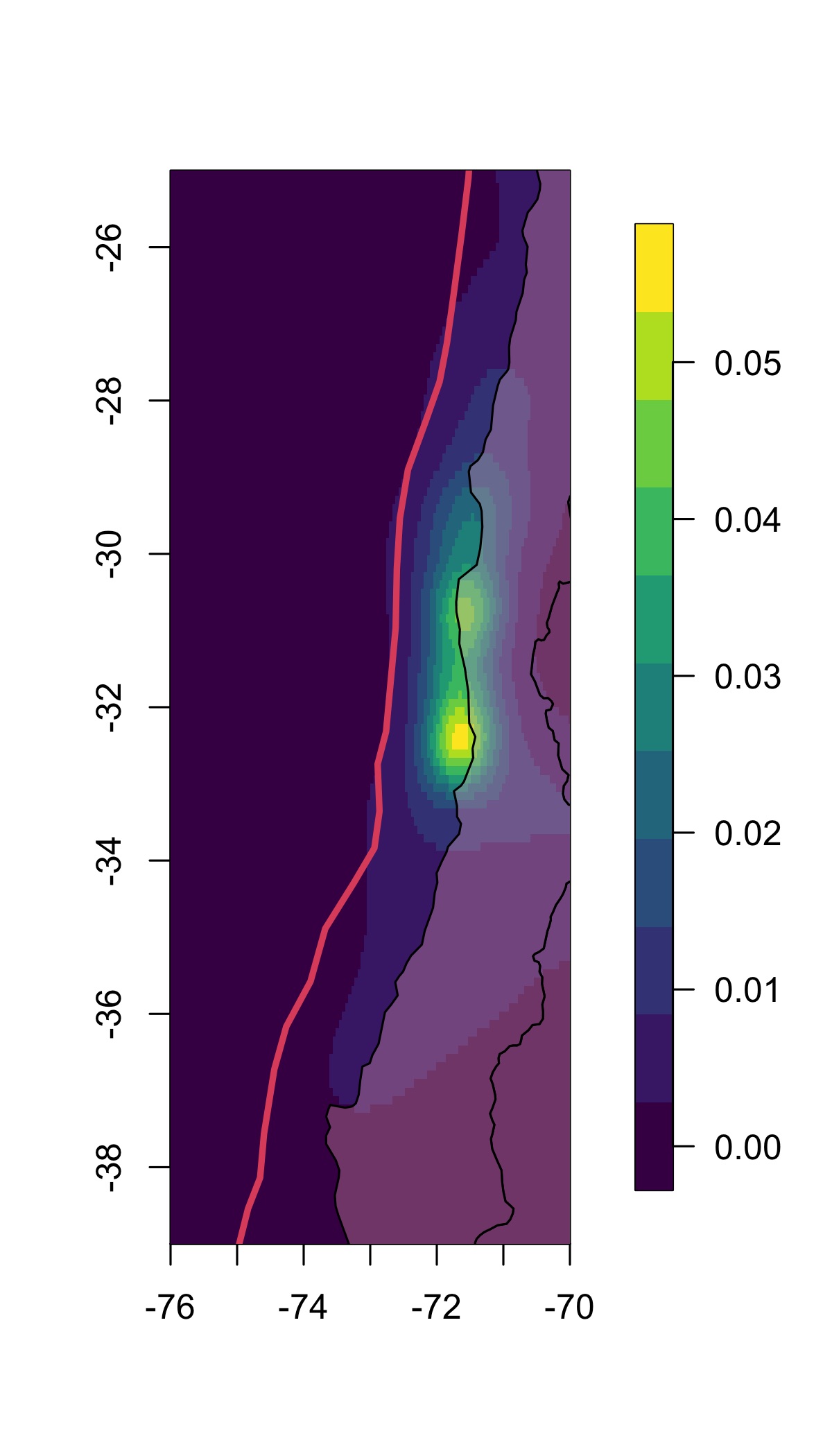}
         \caption{CS-1:1 for Chile A}
         \label{fig:comp_3_mu_13}
     \end{subfigure}
     \hfill
     \begin{subfigure}[b]{0.3\textwidth}
         \centering
         \includegraphics[width=\textwidth]{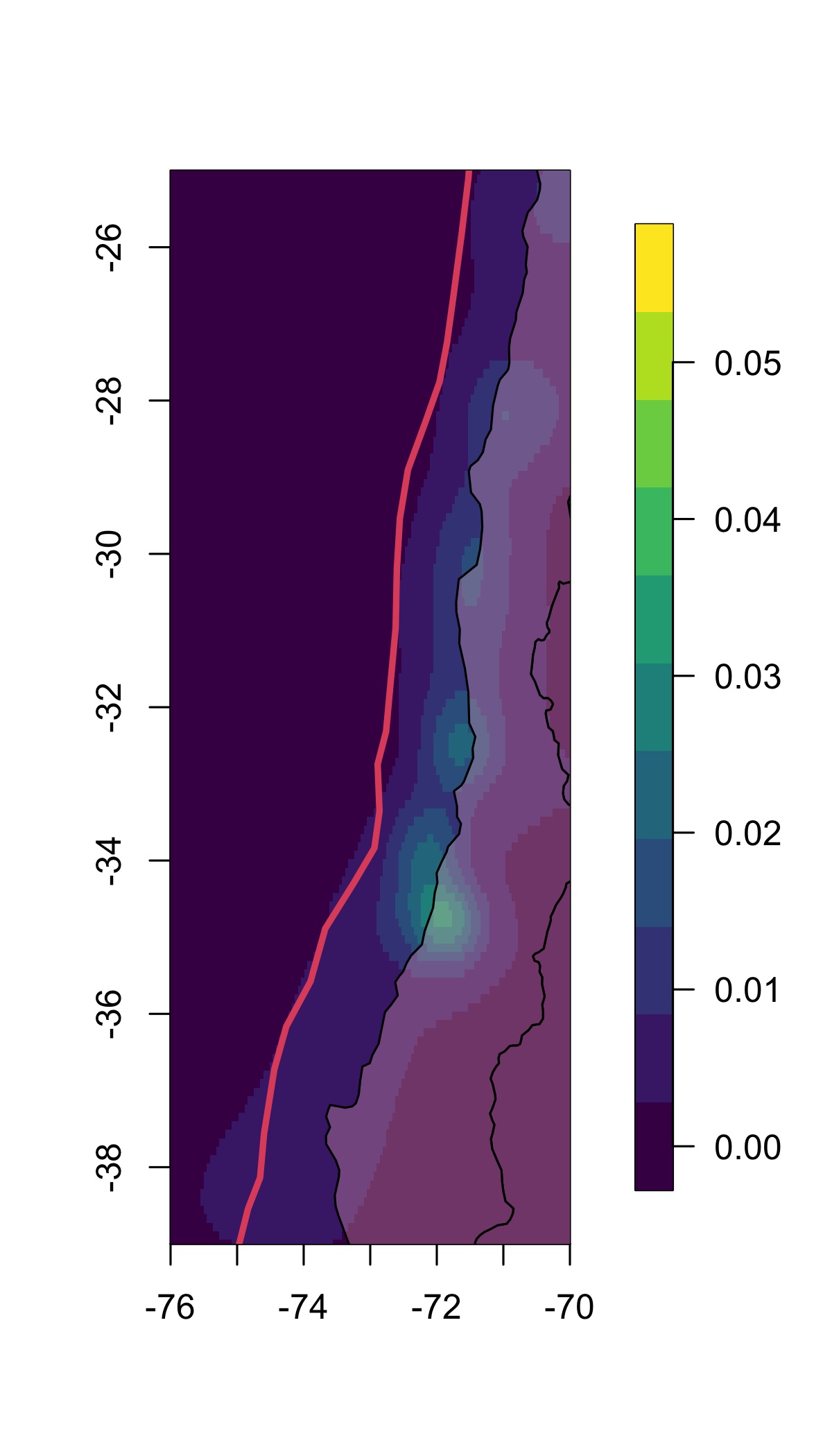}
         \caption{CS-1:1 for Chile B}
         \label{fig:comp_4_mu_13}
     \end{subfigure}
     \hfill
     \begin{subfigure}[b]{0.3\textwidth}
         \centering
         \includegraphics[width=\textwidth]{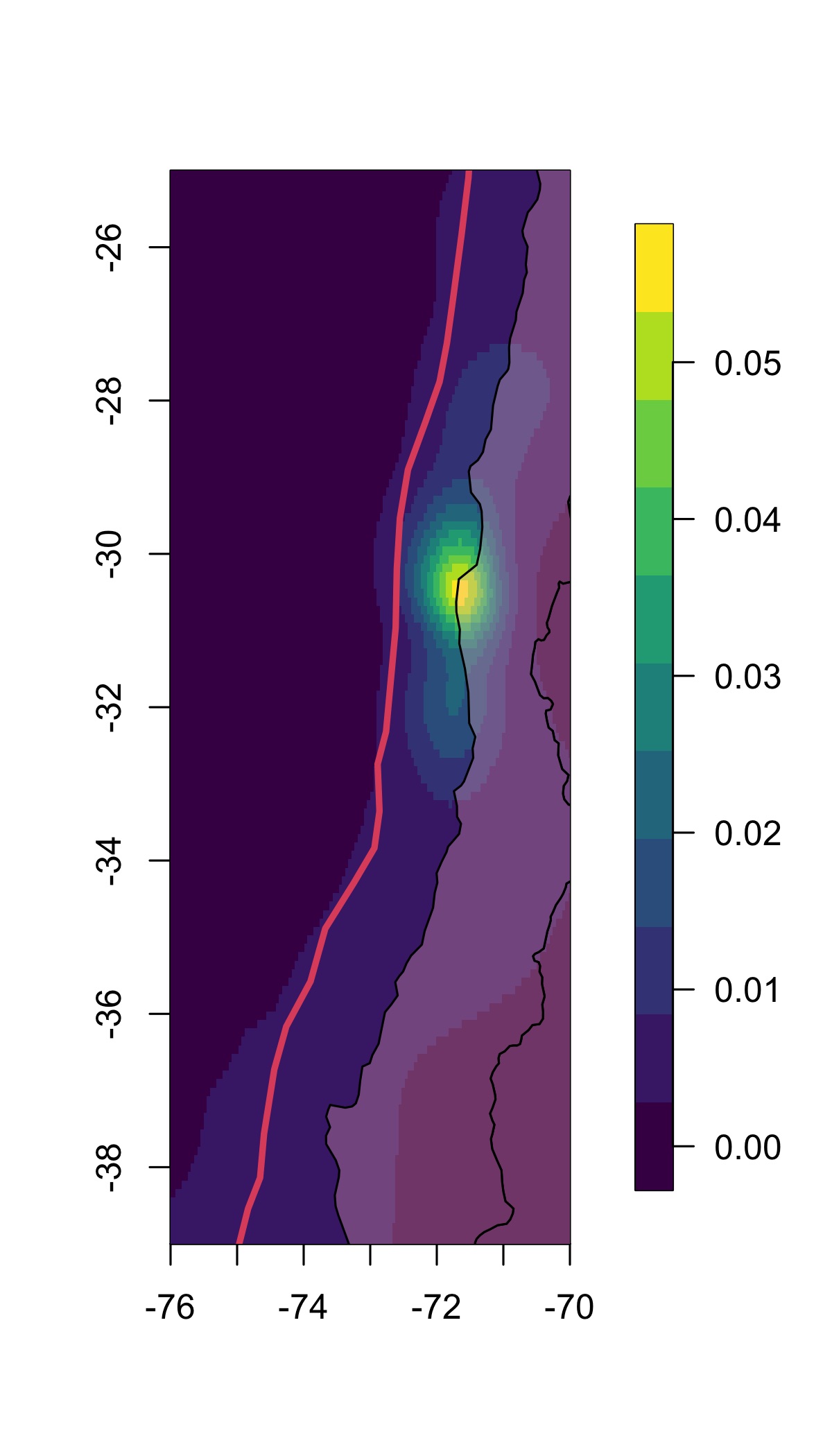}
         \caption{CS-1:1 for Chile C}
         \label{fig:comp_5_mu_13}
     \end{subfigure}
     
     \begin{subfigure}[b]{0.3\textwidth}
         \centering
         \includegraphics[width=\textwidth]{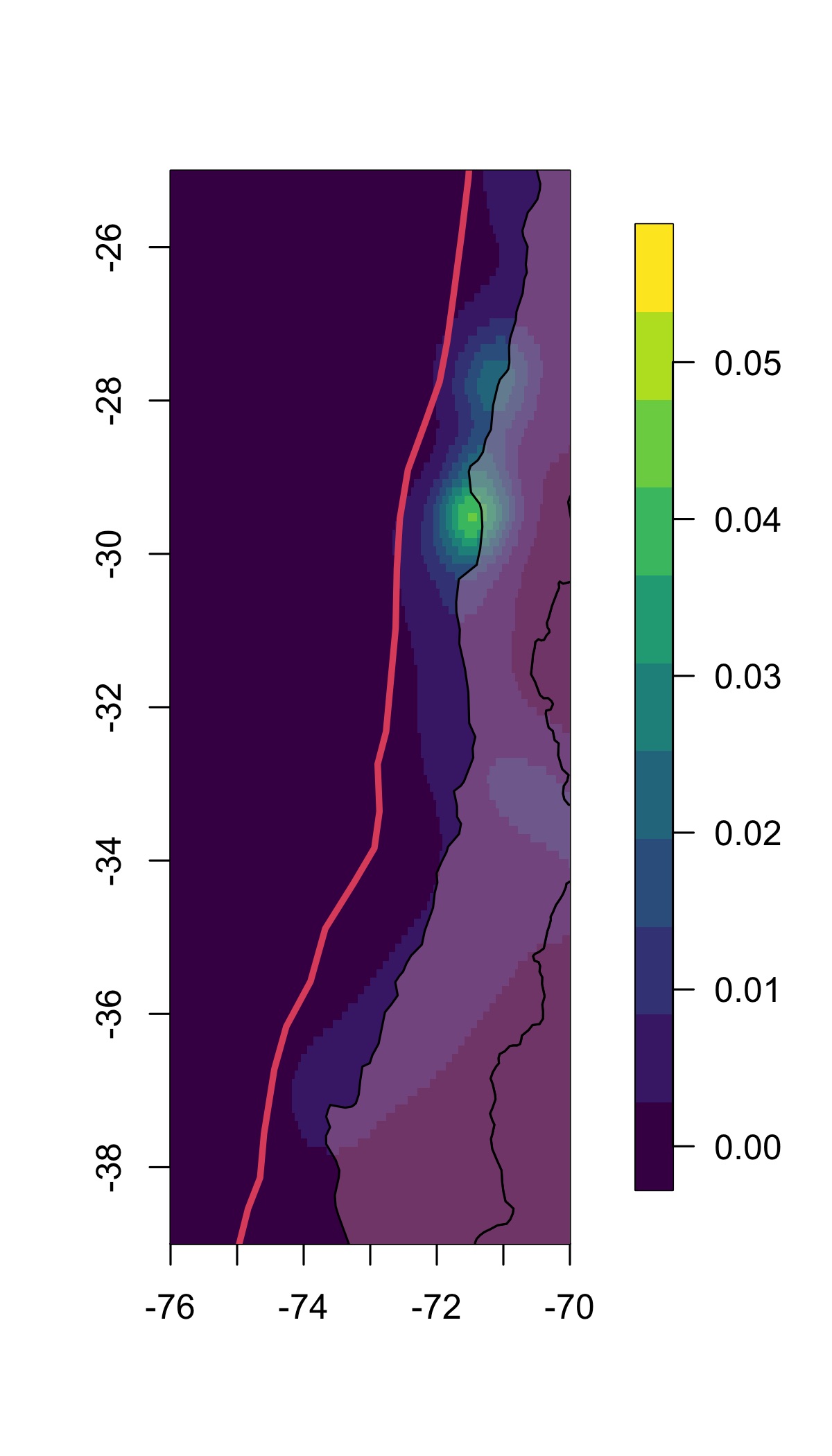}
         \caption{VN-2:1 for Chile A}
         \label{fig:comp_3_mu_2}
     \end{subfigure}
     \hfill
     \begin{subfigure}[b]{0.3\textwidth}
         \centering
         \includegraphics[width=\textwidth]{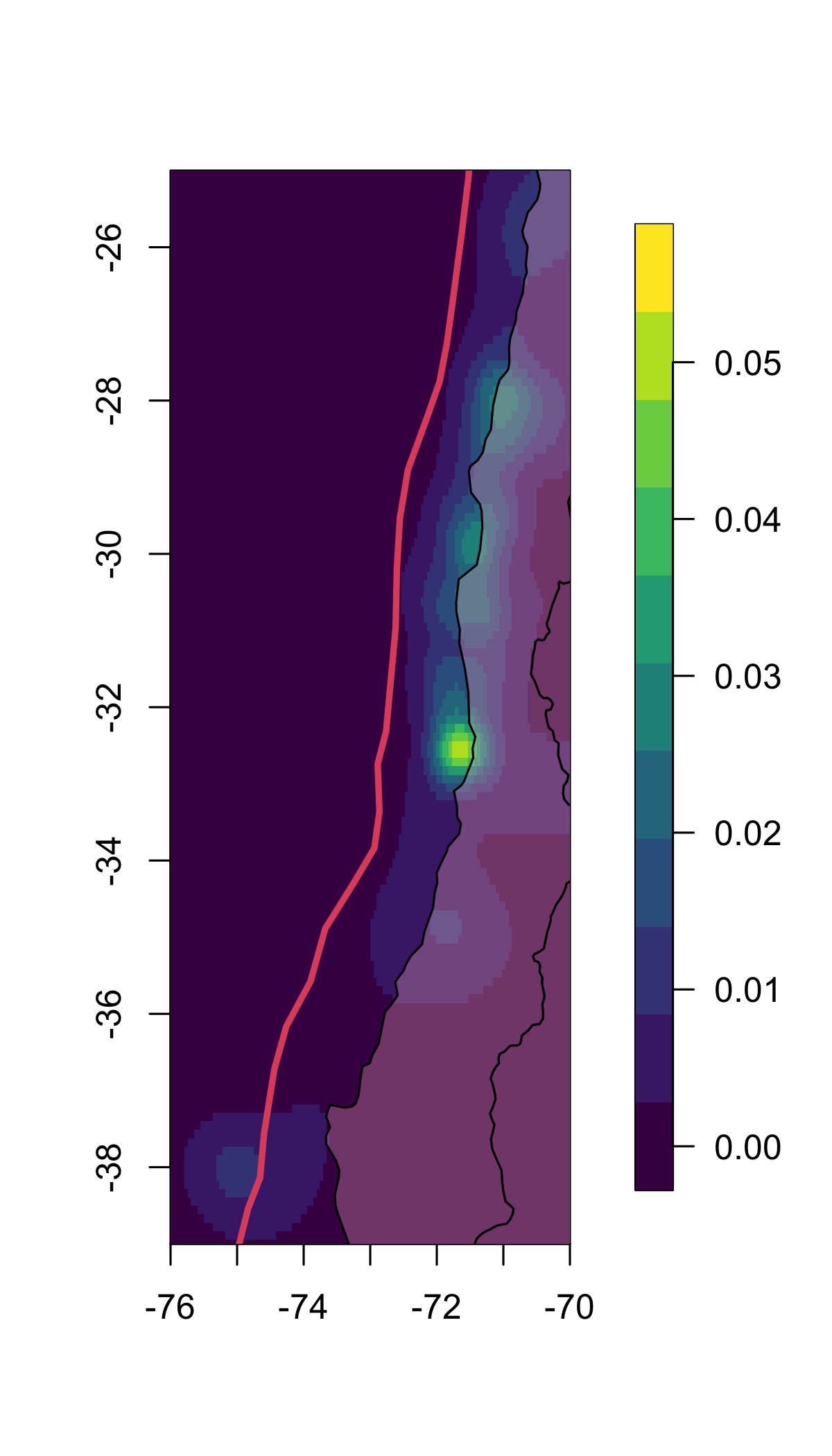}
         \caption{VN-1:1 for Chile B}
         \label{fig:comp_4_mu_1}
     \end{subfigure}
     \hfill
     \begin{subfigure}[b]{0.3\textwidth}
         \centering
         \includegraphics[width=\textwidth]{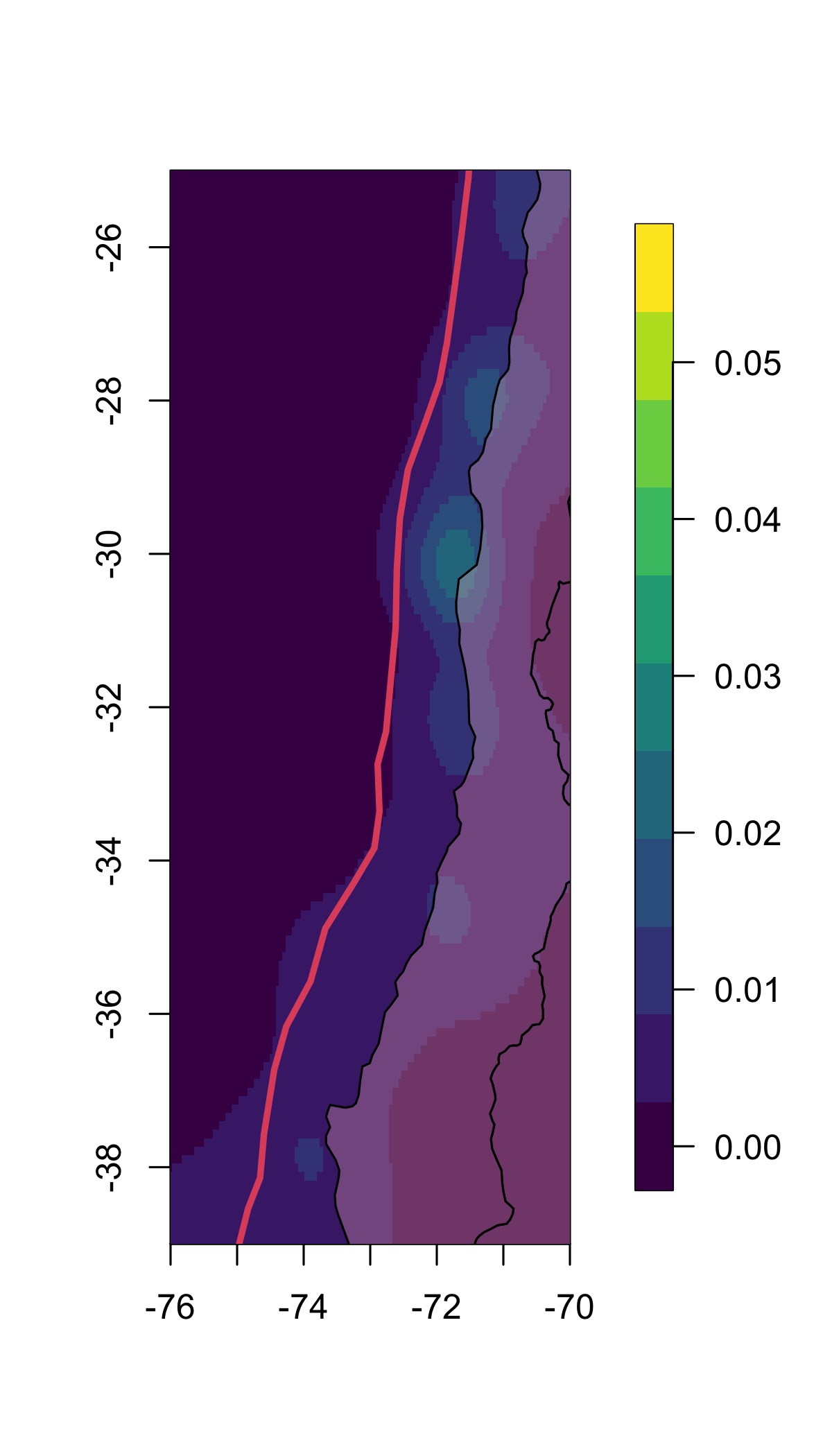}
         \caption{VS-4:1 for Chile C}
         \label{fig:comp_5_mu_8}
     \end{subfigure}
     
     \caption{Comparison of the estimated background rate $\mu(x,y)$ for the three catalogs in the Chile region. The most restrictive model CS-1:1 (top row) is compared to the models with the highest forecast accuracy (bottom row). The solid red line represents the subducting plate boundary.}
     \label{fig:comp_chile}
\end{figure}

\begin{figure}[h!]
     \centering
     \begin{subfigure}[b]{0.45\textwidth}
         \centering
         \includegraphics[width=\textwidth]{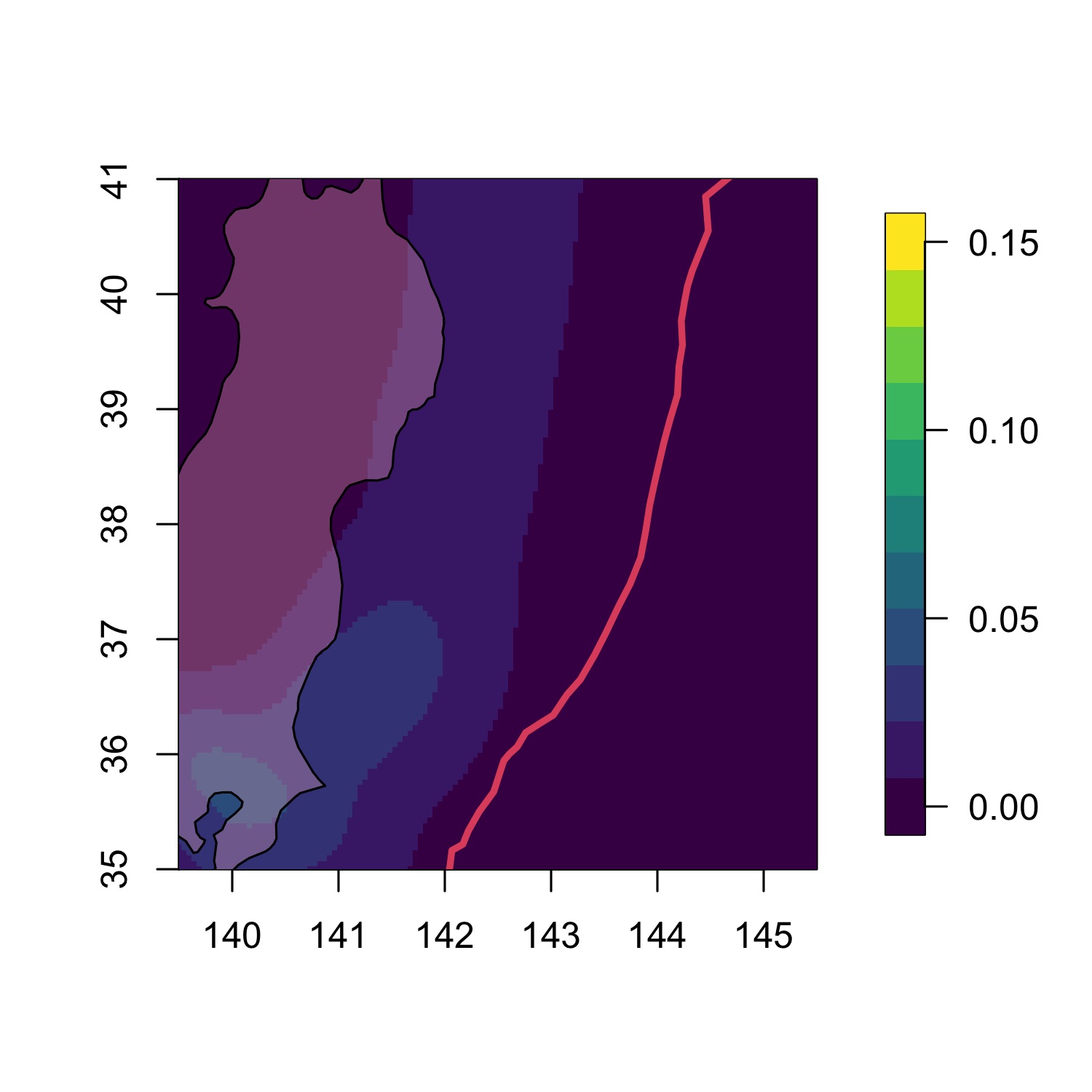}
         \caption{CS-1:1 for Japan A}
         \label{fig:comp_1_mu_13}
     \end{subfigure}
     \hfill
     \begin{subfigure}[b]{0.45\textwidth}
         \centering
         \includegraphics[width=\textwidth]{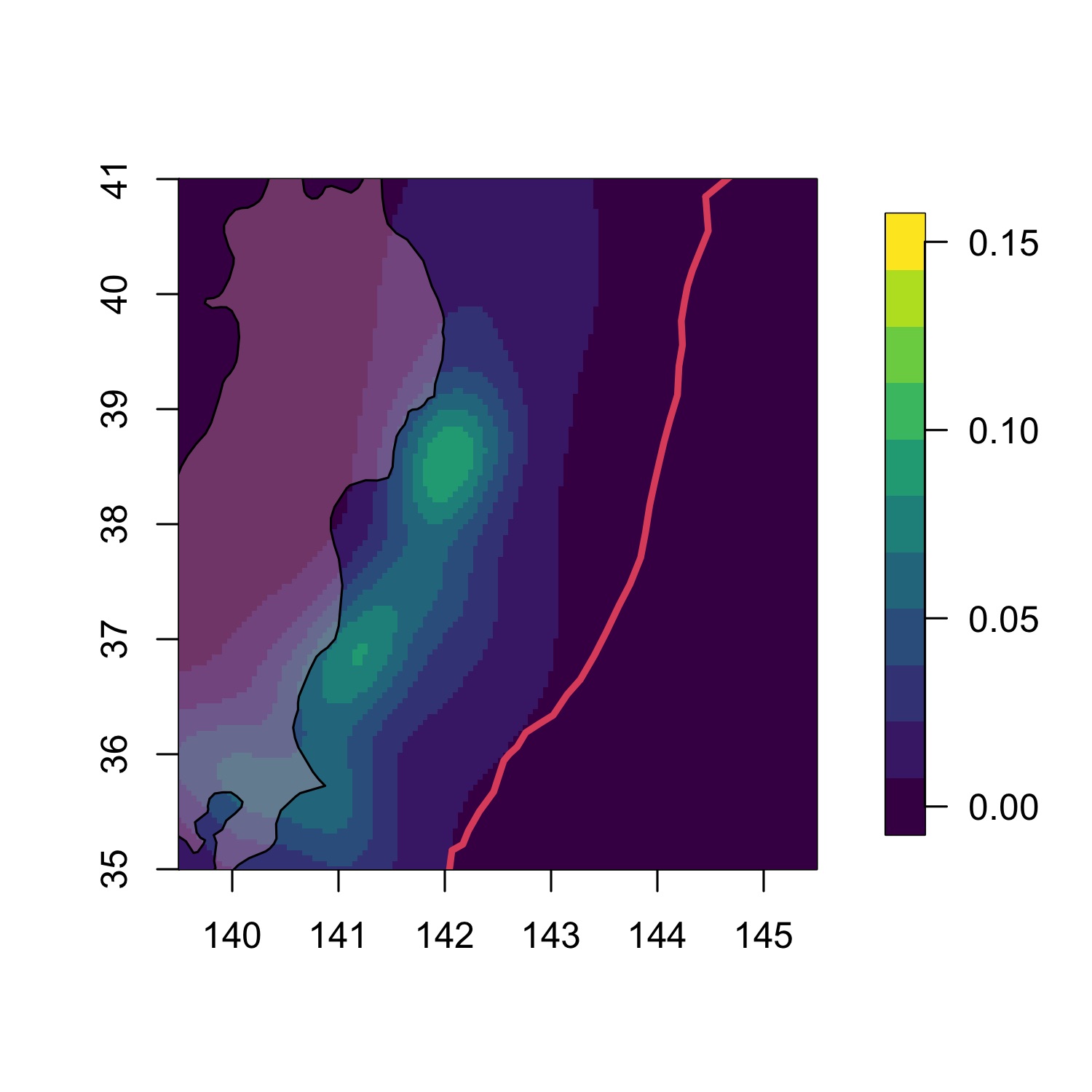}
         \caption{CS-1:1 for Japan B}
         \label{fig:comp_2_mu_13}
     \end{subfigure}
     
     \begin{subfigure}[b]{0.45\textwidth}
         \centering
         \includegraphics[width=\textwidth]{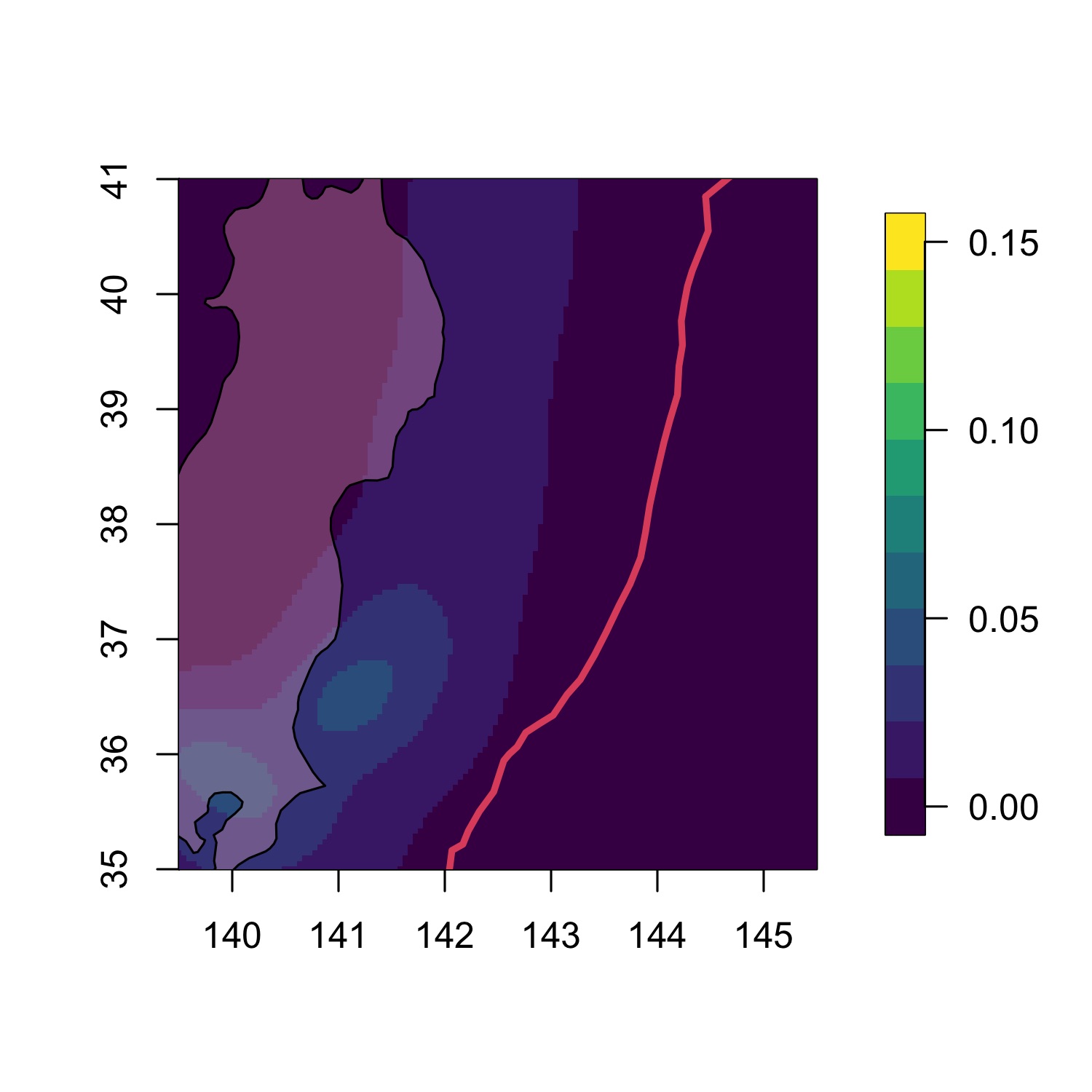}
         \caption{VS-1:1 for Japan A}
         \label{fig:comp_1_mu_5}
     \end{subfigure}
     \hfill
     \begin{subfigure}[b]{0.45\textwidth}
         \centering
         \includegraphics[width=\textwidth]{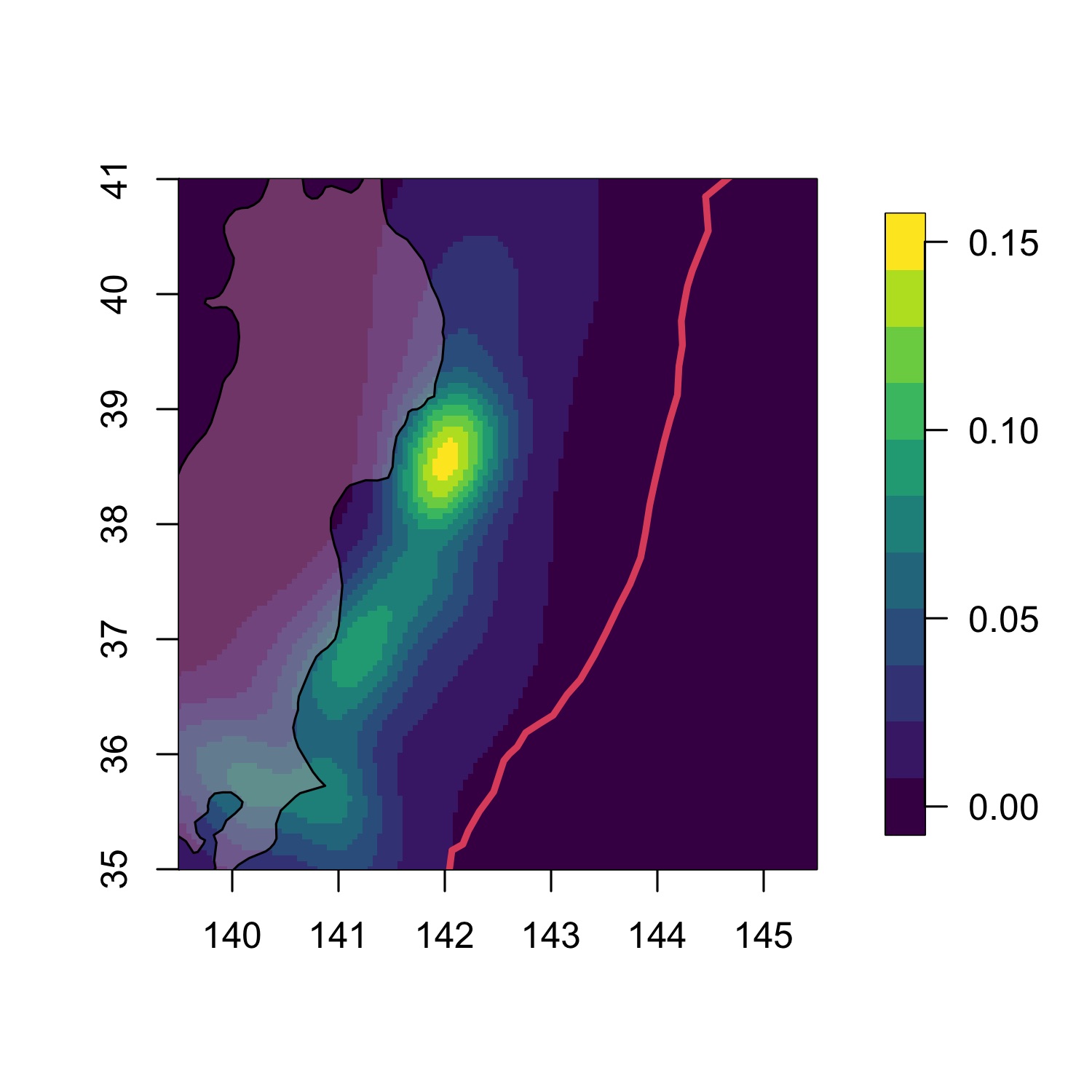}
         \caption{VN-2:1 for Japan B}
         \label{fig:comp_2_mu_1}
     \end{subfigure}
     
     \caption{Comparison of the estimated background rate $\mu(x,y)$ in Japan region. The most restrictive model CS-1:1 (top row) is compared to the models with the highest forecast accuracy (bottom row). Solid red line represents the subducting plate boundary.}
     \label{fig:comp_japan}
\end{figure}

Now we examine the changes in background rates $\mu(x,y)$ before and after the major earthquakes. Figures \ref{fig:comp_chile} and \ref{fig:comp_japan} show the estimated background rates $\mu(x,y)$ for Chile and Japan regions, respectively. Note that we do not need to scale $\mu(x,y)$ with the length of training period, $T$, because of its definition \eqref{eq:background_kernel}. The plots in the top row show the estimation results from the most restrictive model, CS-1:1, while the plots in the bottom row are from the models that provide the best forecast accuracy. When utilizing the CS-1:1 model, the estimated background rate for Chile B is the lowest compared to the other two catalogs in Chile. Allowing model flexibility, on the other hand, has the opposite result, and Chile B becomes the period of the most intense mainshock activity. In the Japan region, using flexible models noticeably increases the background rate for the catalog Japan B while leaving the estimate for Japan A practically unchanged. Another distinctive feature of the flexible models is the change in the overall shape of the background rate distribution, particularly for catalogs from the Chile region. Northward shift is observed for the peaks of the estimated background rate as we allow for spatial variation in aftershock productivity.

\begin{figure}[h!]
     \centering
     \begin{subfigure}[b]{0.3\textwidth}
         \centering
         \includegraphics[width=\textwidth]{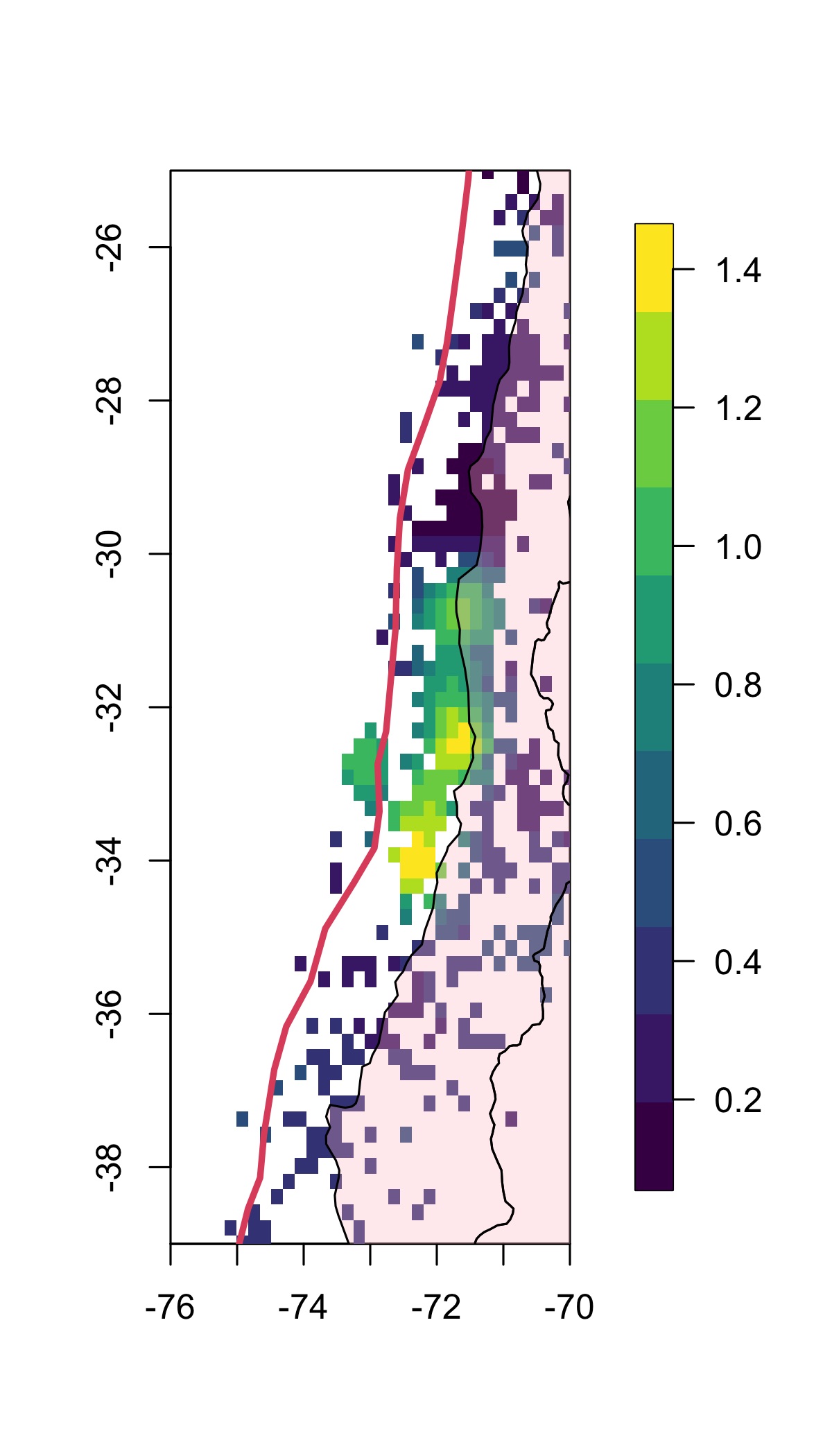}
         \caption{VN-2:1 for Chile A}
         \label{fig:comp_3_alpha_2}
     \end{subfigure}
     \hfill
     \begin{subfigure}[b]{0.3\textwidth}
         \centering
         \includegraphics[width=\textwidth]{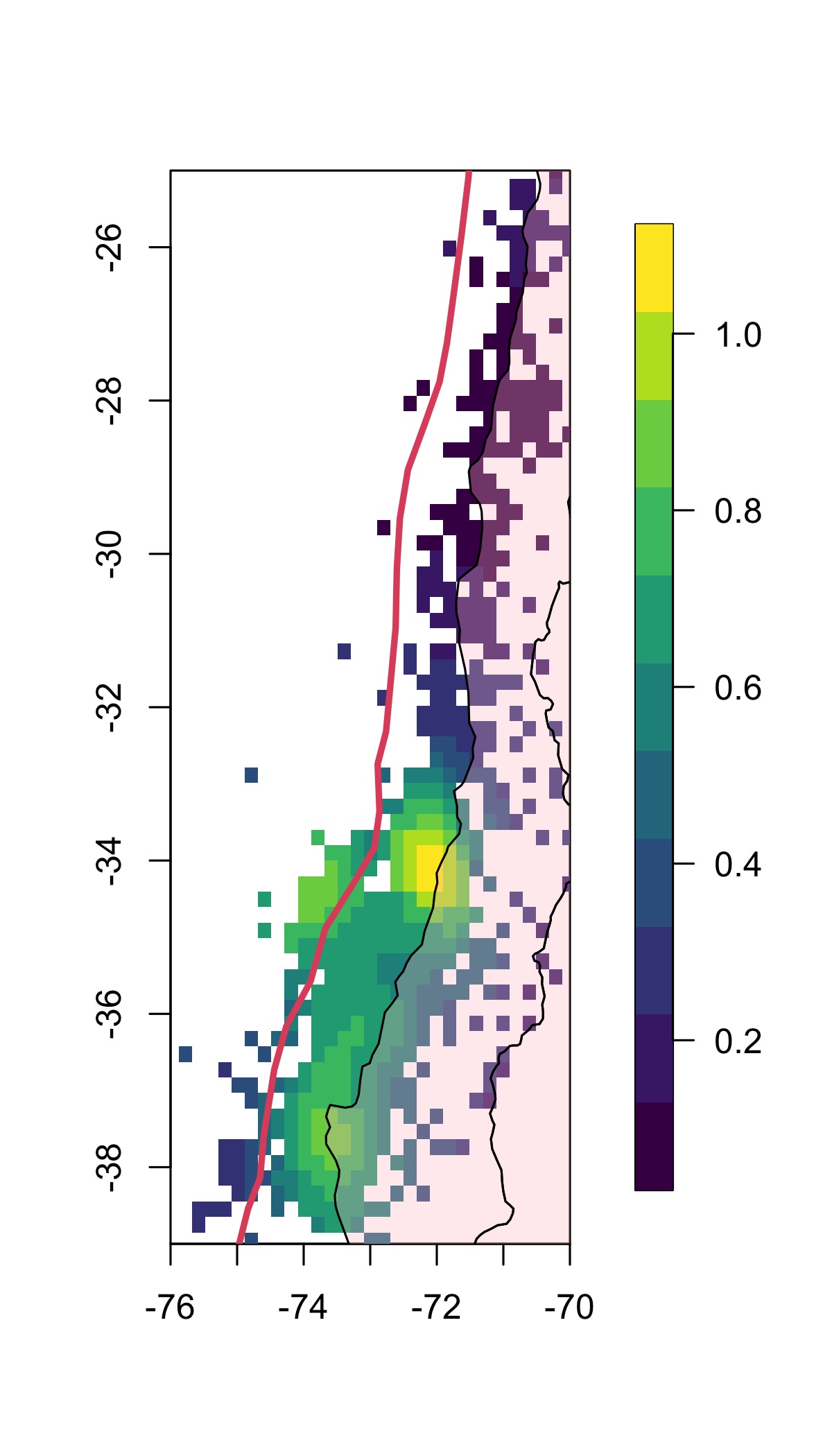}
         \caption{VN-1:1 for Chile B}
         \label{fig:comp_4_alpha_1}
     \end{subfigure}
     \hfill
     \begin{subfigure}[b]{0.3\textwidth}
         \centering
         \includegraphics[width=\textwidth]{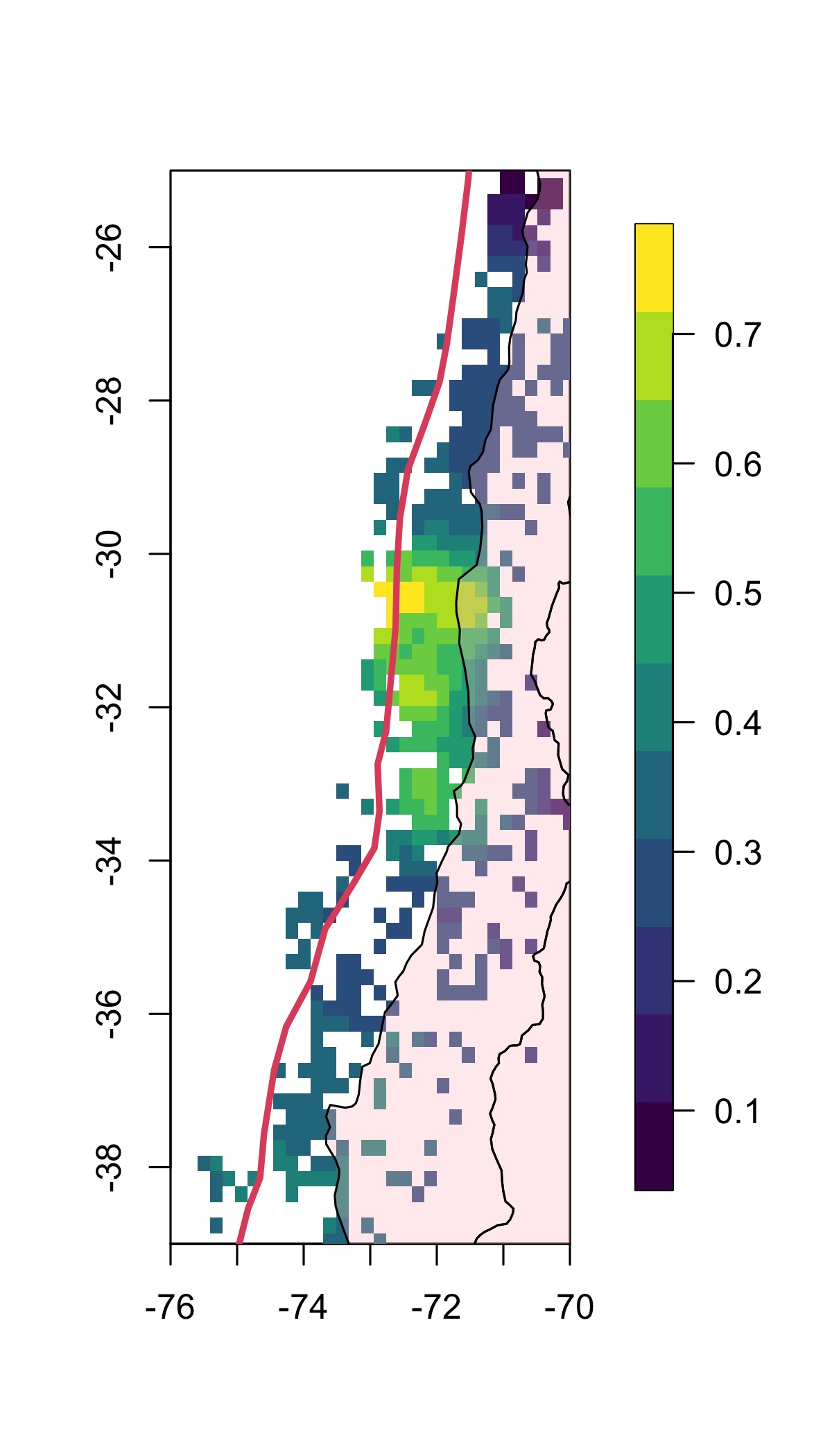}
         \caption{VS-4:1 for Chile C}
         \label{fig:comp_5_alpha_8}
     \end{subfigure}
     
     \caption{\textcolor{black}{Comparison of the spatially-varying aftershock productivity at the cutoff magnitude 4.0. Each output is based on the estimation result from the best model in each period of Chile region.} Solid red line represents the subducting plate boundary.}
     \label{fig:comp_chile_alpha}
\end{figure}

\begin{figure}[h!]
     \centering
     \begin{subfigure}[b]{0.45\textwidth}
         \centering
         \includegraphics[width=\textwidth]{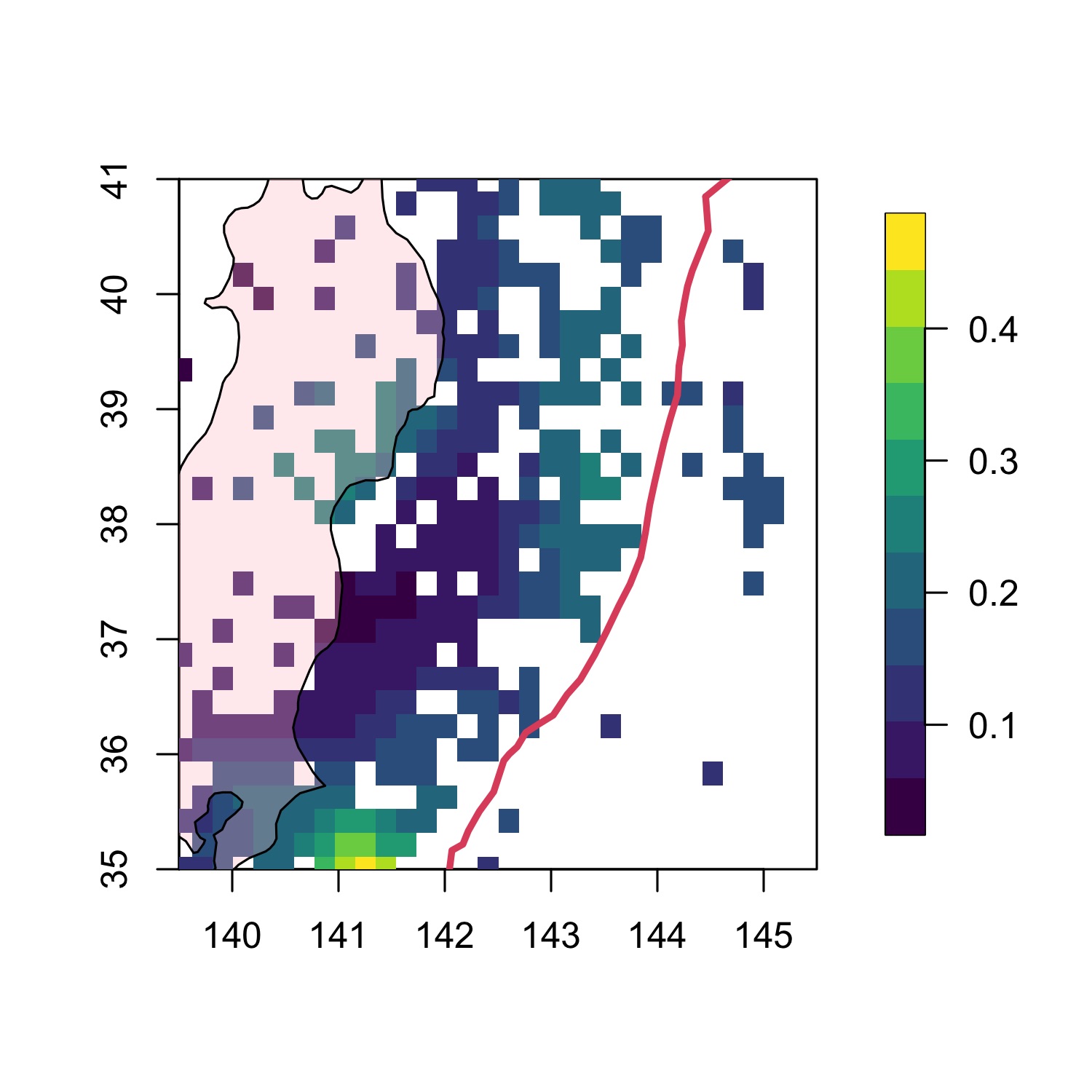}
         \caption{VS-1:1 for Japan A}
         \label{fig:comp_1_alpha_5}
     \end{subfigure}
     \hfill
     \begin{subfigure}[b]{0.45\textwidth}
         \centering
         \includegraphics[width=\textwidth]{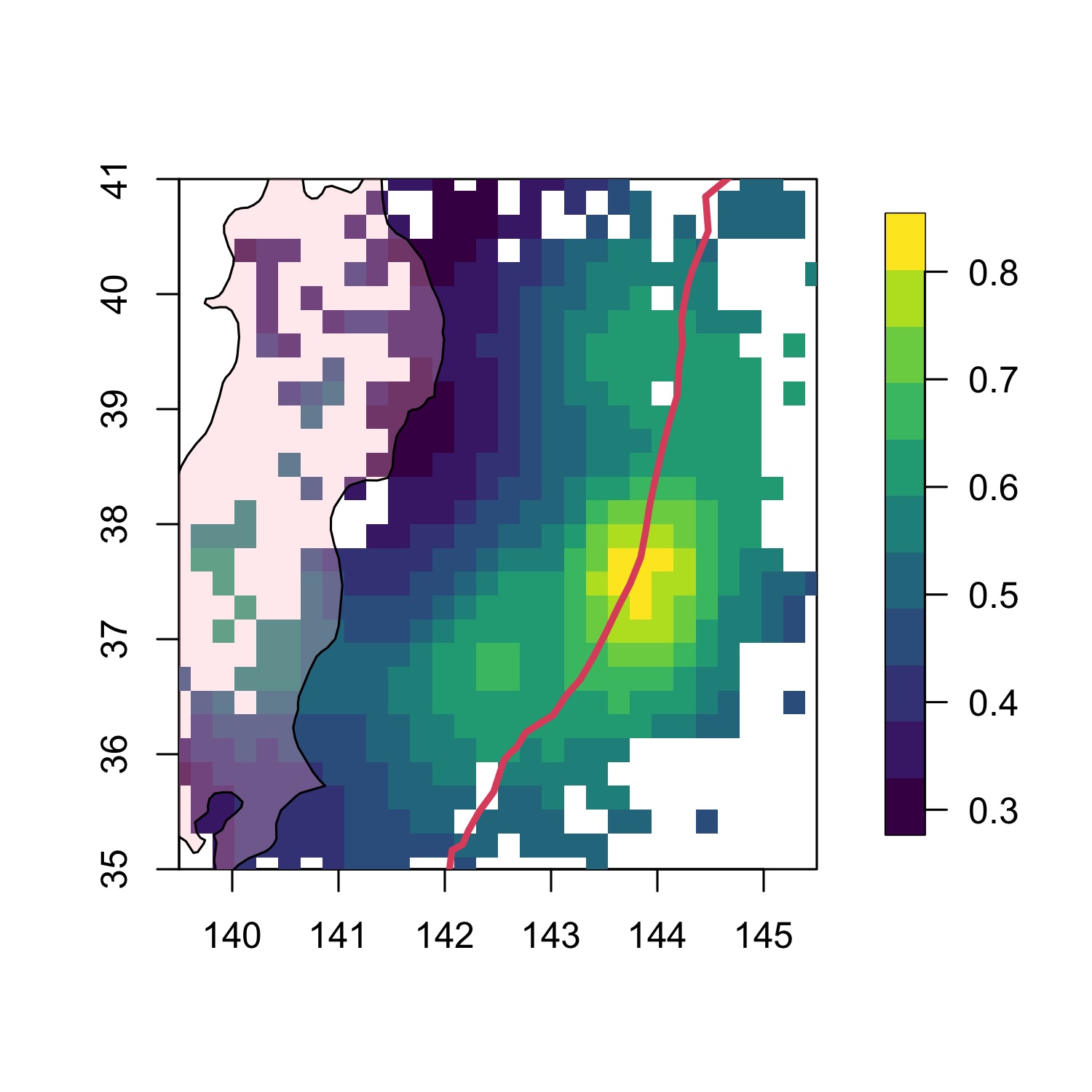}
         \caption{VN-1:1 for Japan B}
         \label{fig:comp_2_alpha_1}
     \end{subfigure}

     \caption{\textcolor{black}{Comparison of the spatially-varying aftershock productivity at the cutoff magnitude 4.0. Each output is based on the estimation result from the best model in each period of Japan region.} Solid red line represents the subducting plate boundary.}
     \label{fig:comp_japan_alpha}
\end{figure}

One of the reasons which attribute to these phenomena is spatially varying productivity, which is assumed in all the flexible models with the highest forecast accuracy. \textcolor{black}{Figures \ref{fig:comp_chile_alpha} and \ref{fig:comp_japan_alpha} illustrate the estimates of aftershock productivities for the earthquakes at the cutoff magnitude, $\alpha(x,y) \kappa(4.0)$, for the best model of each catalog in Chile and Japan, respectively. By observing these figures, we can identify the region with active aftershock occurrences compared to other region.} So, the estimated background rate becomes lower in the corresponding area as we assume the spatial variability in aftershock productivity. \textcolor{black}{On the contrary, a region of low aftershock productivity yields relatively more mainshocks compared to the models that assume $\alpha(x,y) = 1$. Regarding the catalogs of our interest, Chile B has a dramatic change in the aftershock productivity at latitude $L = -33^\circ$ (Figure \ref{fig:comp_4_alpha_1}) making the corresponding period have vigorous mainshock activity, and Japan B shows its lowest aftershock productivity along the coast around the latitude $L = 39^\circ$ (Figure \ref{fig:comp_2_alpha_1}) making the original peak of background rate even higher. On the other hand, catalogs Chile A and C have high aftershock productivity between the latitudes $L = -30^\circ$ and $L = -34^\circ$ while Chile B has high values between $L = -33^\circ$ and $L = -39^\circ$.} These portions of spatial domain coincides with the ones where the estimated background rates from CS-1:1 model become lower as we allow model flexibility for better forecast. Finally, catalog Japan A has $\alpha(x,y) \approx 1$ on the portion of spatial domain where there are many earthquake occurrences (Figure \ref{fig:comp_1_alpha_5}), and this can be a reason why we cannot achieve a significant improvement in forecast accuracy via spatially varying aftershock productivity.

\textcolor{black}{Note that we do not present the estimated aftershock productivity at the cutoff magnitude for the location where there is no earthquakes nearby in Figures \ref{fig:comp_chile_alpha} and \ref{fig:comp_japan_alpha}.} Although $\alpha(x,y)$ can be obtained for every point in the spatial domain, its definition \eqref{eq:alpha_star_est} as a ratio of two local averages can give unstable and misleading results for the portion where there are little observation. Instead, we divide the spatial domain into $0.2^\circ \times 0.2^\circ$ cells and use the average of the estimated $\alpha(x,y)$ for better visual representation.

\section{Discussion}\label{sec:5}
We propose a new spatio-temporal flexible Hawkes model on earthquake occurrences which builds on the previous works on ETAS models to focus on understanding the aftershock dynamics. To the best of our knowledge, this is the first attempt to use nonparametric ETAS models to allow aftershock productivity to vary based on the spatial location as well as magnitude. We achieve further flexibility by considering seismicity anisotropy and space-time interaction (via non-separable structure) in aftershock occurrences. All of these new properties are incorporated into the fully kernel-based ETAS model, in which we have extended the histogram-based ones to obtain smoothly varying estimates. \textcolor{black}{Stability of the model estimation is demonstrated in the Appendix B by using the synthetic earthquake catalogs simulated from a parametric ETAS model with inhomogeneous background rate and spatially constant aftershock productivity, $\alpha(x,y) = 1$. The results confirm that spatially varying 
$\alpha$-function (thus more flexible than true) is not causing instability in terms of the model estimation and forecast accuracy.} By applying various combinations of the proposed approaches to earthquake data from Chile and Japan, we have demonstrated improved forecast accuracy. We have also investigated possible new explanations for the change in mainshock activity before and after major earthquakes.

\textcolor{black}{The research presented in this study is a step towards our goal to build an earthquake forecast model that reflects the nature of their occurrences more flexibly. We plan to test our eventual model in the CSEP (Collaboratory for the Study of Earthquake Predictability) to see how it compares to other forecast models and to look into the prospect of further progress.} The kernel bandwidth selection is one of the challenges that our model encounters, which is inherent for nonparametric models in general. Cross validation is a commonly adopted solution to select the kernel bandwidth. However, the proposed model \eqref{eq:proposed_conditional_intensity} has four components that needs bandwidth selection: $\mu$, $\alpha$, $\kappa$, and $g$.  Furthermore, iterative estimation algorithm changes the weights in the kernel estimators for every iteration. This implies that the bandwidth selection needs to be conducted more than once. Though we set them to appropriate values for all catalogs and focus on changing the features of interest, an objective method for bandwidth selection needs to be established for more precise analysis.

\textcolor{black}{On the other hand, we plan to allow the triggering density $g(\Delta x, \Delta y, \Delta t; \eta, \theta)$ to have spatially varying anisotropy parameters (also including the depth direction beyond the 2--D seismicity considered here) as future research.} Approximating the fault plane orientation with a plate boundary is a crude and oversimplified approach. Depending on the spatial domain of a given catalog, there may be a number of faults that are not parallel to the plate boundary. Fortunately, the fault plane (despite an ambiguous auxiliary plane) can be estimated based on the first motion of seismic waves. The Global Centroid-Moment-Tensor (GlobalCMT) Project inverts and provides the fault-plane solutions, available online at their website \citep{dziewonski1981determination, ekstrom2012global}. We may be able to build a more accurate ETAS model if we could model $\eta$ and $\theta$ based on the seismologically estimated strike, dip, and slip, as well as the associated stress change at the aftershock location due to the mainshocks \citep{Hill2009,toda2012aftershocks}. However, only earthquakes with magnitudes 5.0 or greater are available in GlobalCMT. 

Other than the topics mentioned above, future research areas for the nonparmetric ETAS models include modeling of the earthquake occurrences considering their focal depths. We can also develop a nonparametric ETAS model which accounts for the Utsu-Seki law by scaling the spatial lags. The law says that the spatial range of aftershocks is related with the mainshock magnitude in an exponential fashion.

\newpage

\section*{Declaration of Interest}
None

\section*{Acknowledgements}
Mikyoung Jun acknowledges support by NSF DMS-1925119 and DMS-2123247.

\bibnote{us2017advanced}{[dataset]}
\bibnote{HugoAhlenius}{[dataset]}
\bibliographystyle{elsarticle-harv}

\bibliography{references.bib}

\newpage
\bigskip
\begin{center}
{\large\bf Appendix A: Modified MISD algorithm }
\end{center}

\begin{itemize}
    \item[1.] Initialize the triggering probability matrix $P^{(0)} = ( p^{(0)}_{ij} )_{1\leq i,j \leq N}$ as
$$p^{(0)}_{ij} = 1/i \text{ for } j = 1, 2, \cdots, i.$$
    \item[2.] Iterate for $\ell = 1, 2, \cdots$, until $\max_{i,j} |p^{(\ell)}_{ij} - p^{(\ell-1)}_{ij}|$ becomes smaller than the convergence criterion $\epsilon > 0$.
    \begin{itemize}
        \item[a.] Estimate the background rate
 $$\mu^{(\ell)}(x,y) \leftarrow \frac{1}{q_{h_1}(x,y|D) \cdot T} \sum_{i=1}^N p^{(\ell-1)}_{ii} G_{h_1}(x - x_i, y - y_i),$$
        where $T$ is a length of the temporal domain, $D$ is a spatial domain, and $q_{h_1}(x,y|D)$ is an edge-correction factor at $(x,y)$.
 
        \item[b.]  Estimate the productivity function
 $$\kappa^{(\ell)}(m) \leftarrow \frac{\sum_{j=1}^{N-1} (\sum_{i=j+1}^N p_{ij}^{(\ell-1)}) G_{h_2}(m - m_j)}{\sum_{j=1}^{N-1} 
 G_{h_2}(m - m_j)}.$$
 
        \item[c.] Estimate the regional productivity correction factor
 $$\alpha^{(\ell)}(x,y) \leftarrow \frac{1}{A^{*}} \frac{\sum_{j=1}^{N-1} (\sum_{i=j+1}^N p^{(\ell-1)}_{ij}) G_{h_3}(x - x_j, y - y_j)}{\sum_{j=1}^{N-1} \kappa^{(\ell)}(m_j) G_{h_3}(x - x_j, y - y_j)},$$
        where $A^* = \sum_{j=1}^{N-1} \sum_{i=j+1}^N p^{(\ell-1)}_{ij}/\sum_{j=1}^{N-1} \kappa^{(\ell)}(m_j)$.
 
        \item[d.] Estimate the triggering density
        \begin{equation*}
        \begin{split}
            g_0^{*(\ell)}(\Delta \mathbf{s}^*,\Delta t^*) &\leftarrow \frac{\sum_{i>j} p^{(\ell-1)}_{ij} G_{h_4} (\Delta \mathbf{s}^* - \Delta \mathbf{s}^*_{ij}, \Delta t^* - \Delta t^*_{ij})}{q_{h_4}(\Delta \mathbf{s}^*, \Delta t^*|\mathbb{R}^{+} \times \mathbb{R}^{+}) \sum_{i>j} p^{(\ell-1)}_{ij} \sum_{i>j}1},\\
            g_0^{(\ell)}(\Delta \mathbf{s},\Delta t) &\leftarrow \frac{g_0^{*(\ell)}(\Delta \mathbf{s}^*,\Delta t^*)}{\sigma_s \sigma_t \exp(\sigma_s \Delta \mathbf{s}^* + \sigma_t \Delta t^*)},
        \end{split}
        \end{equation*}
        where $\Delta \mathbf{s}^*_{ij} = \log(\Delta \mathbf{s}_{ij} + 1) / \sigma_s, \ \Delta t^*_{ij} = \log(\Delta t_{ij}+1) / \sigma_t$, $\Delta \mathbf{s}_{ij} = \sqrt{(x_i - x_j)^2 + (y_i - y_j)^2}$, $\Delta t_{ij} = t_i - t_j$, $\sigma_s$ is a standard deviation of $\log(\Delta \mathbf{s}_{ij} + 1)$, $\sigma_t$ is a standard deviation of $\log(\Delta t_{ij} + 1)$, and $q_{h_4}(\Delta \mathbf{s}^*, \Delta t^*|\mathbb{R}^{+} \times \mathbb{R}^{+})$ is an edge-correction factor at $(\Delta \mathbf{s}^*, \Delta t^*)$.

        \item[e.]  Update the triggering probability matrix $P^{(\ell)}$ as
        \begin{equation*}
        \begin{split}
        p_{ii}^{(\ell)} &\leftarrow \frac{\mu^{(\ell)}(x_i,y_i)}{\mu^{(\ell)}(x_i,y_i) + \sum_{j: t_j < t_i}\alpha^{(\ell)}(x_j, y_j) \kappa^{(\ell)}(m_j) g_0^{(\ell)}(\Delta \mathbf{s}_{ij}, \Delta t_{ij}) / (2\pi \Delta \mathbf{s}_{ij})},\\
        p_{ij}^{(\ell)} &\leftarrow \frac{\alpha^{(\ell)}(x_j, y_j) \kappa^{(\ell)}(m_j) g_0^{(\ell)}(\Delta \mathbf{s}_{ij}, \Delta t_{ij}) / (2\pi \Delta \mathbf{s}_{ij})}{\mu^{(\ell)}(x_i,y_i) + \sum_{j: t_j < t_i}\alpha^{(\ell)}(x_j, y_j) \kappa^{(\ell)}(m_j) g_0^{(\ell)}(\Delta \mathbf{s}_{ij},\Delta t_{ij}) / (2\pi \Delta \mathbf{s}_{ij})}.
        \end{split}
        \end{equation*}
    \end{itemize}
\end{itemize}

\newpage
\bigskip
\begin{center}
\textcolor{black}{{\large\bf Appendix B: Stability of the model estimation}}
\end{center}
We illustrate the stability of the proposed model by using synthetic earthquake data. We generate a synthetic earthquake catalog from a parametric ETAS model with an inhomogeneous background rate and a spatially constant aftershock productivity function, i.e. $\alpha(x,y) = 1$. The forecast accuracy and the estimated background rate are then compared between two models, VS-1:1 and CS-1:1, which share the assumptions of spatial isotropy and space-time separability in the aftershock occurrences but differ in whether $\alpha(x,y)$ varies over the spatial domain or not.

For the simulation, we produce 200 synthetic earthquake catalogs over a square shape spatial domain $\{(x,y): 0^\circ \leq x, y \leq 6^\circ \}$ for 4400 days utilizing the branching structure of the Hawkes processes \citep{zhuang2004analyzing}. For each catalog, we discard the observations in the first 2000 days to use the data in a steady state. The observations from the next 2000 days are then used to fit the ETAS models, while the remaining observations from the last 400 days are used to measure forecast accuracy. Assumed parametric ETAS model for the data generation has the form
\begin{equation*}
    \lambda(x,y,t|\mathcal{H}_t) = \mu(x,y) + \sum_{\{j:t_j < t\}} \kappa(m_j) g_1(x - x_j, y - y_j) g_2(t - t_j),
\end{equation*}
where $\mu(x,y) = 0.0125 \cdot I_{\{(x,y):1\leq x,y \leq 5\}}(x,y) + 0.05 \cdot I_{\{(x,y): 3\leq x \leq 5, 1 \leq y \leq 5\}}(x,y)$, $\kappa(m) =  0.2  \exp(1.7 (m - 4) )$, $g_1(\Delta x, \Delta y) = \frac{0.668}{0.00204 \pi} (1 + \frac{\Delta x^2 + \Delta y^2}{0.00204})^{-1.668}$, and $g_2(\Delta t) = \frac{0.0947}{0.0327} (1 + \frac{\Delta t}{0.0327})^{-1.0947}$. Note that the background rate is nonzero only in the middle of the spatial domain to reduce the edge effect in the estimation, and the spatial and temporal triggering densities are using the estimation results in \cite{ogata1998space} as their parameters.  In terms of magnitude distribution, we suppose that it is determined independently of past occurrences using an exponential distribution with rate $\log 10$, i.e. $J(m) \sim 4 + \mbox{Exp}(\log 10)$.

\begin{figure}[h!]
     \centering
     \begin{subfigure}[b]{0.49\textwidth}
         \centering
         \includegraphics[width=\textwidth]{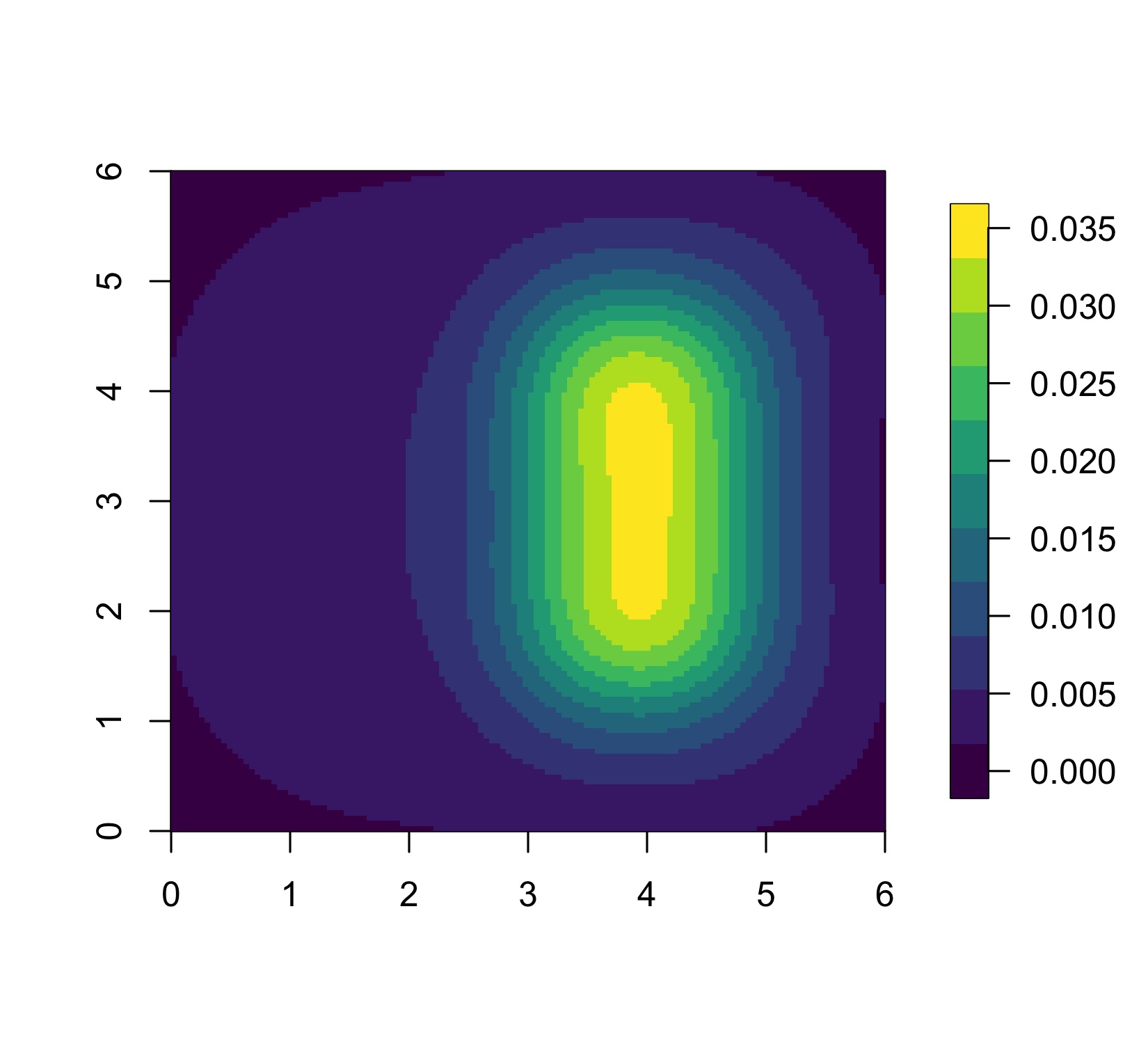}
         \label{fig:synthetic_bgr_mean}
         \caption{VS:1-1, Pixelwise average}
     \end{subfigure}
     \hfill
     \begin{subfigure}[b]{0.49\textwidth}
         \centering
         \includegraphics[width=\textwidth]{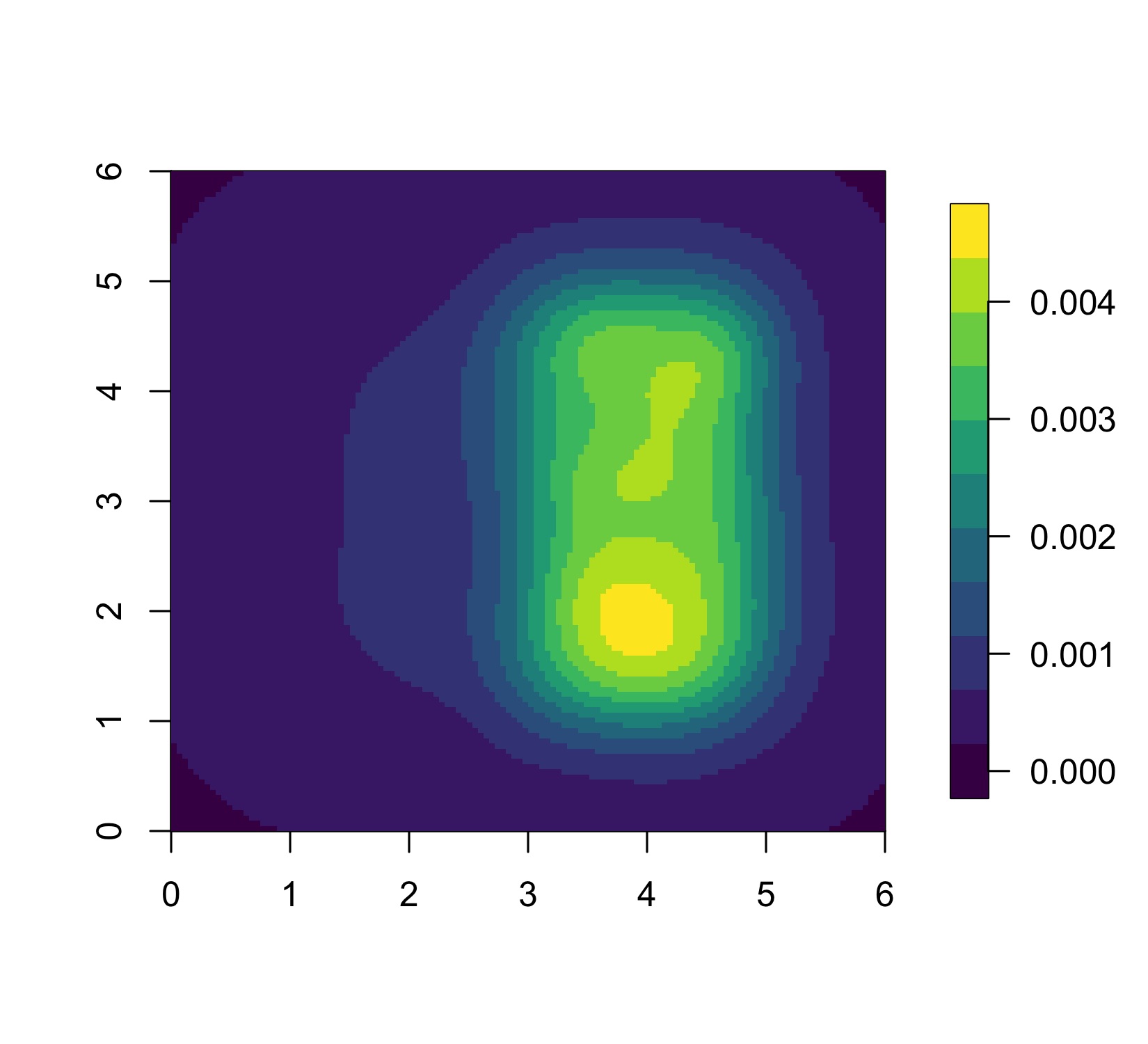}
         \label{fig:synthetic_bgr_sd}
         \caption{VS:1-1, Pixelwise standard deviation}
     \end{subfigure}\\
     \begin{subfigure}[b]{0.49\textwidth}
         \centering
         \includegraphics[width=\textwidth]{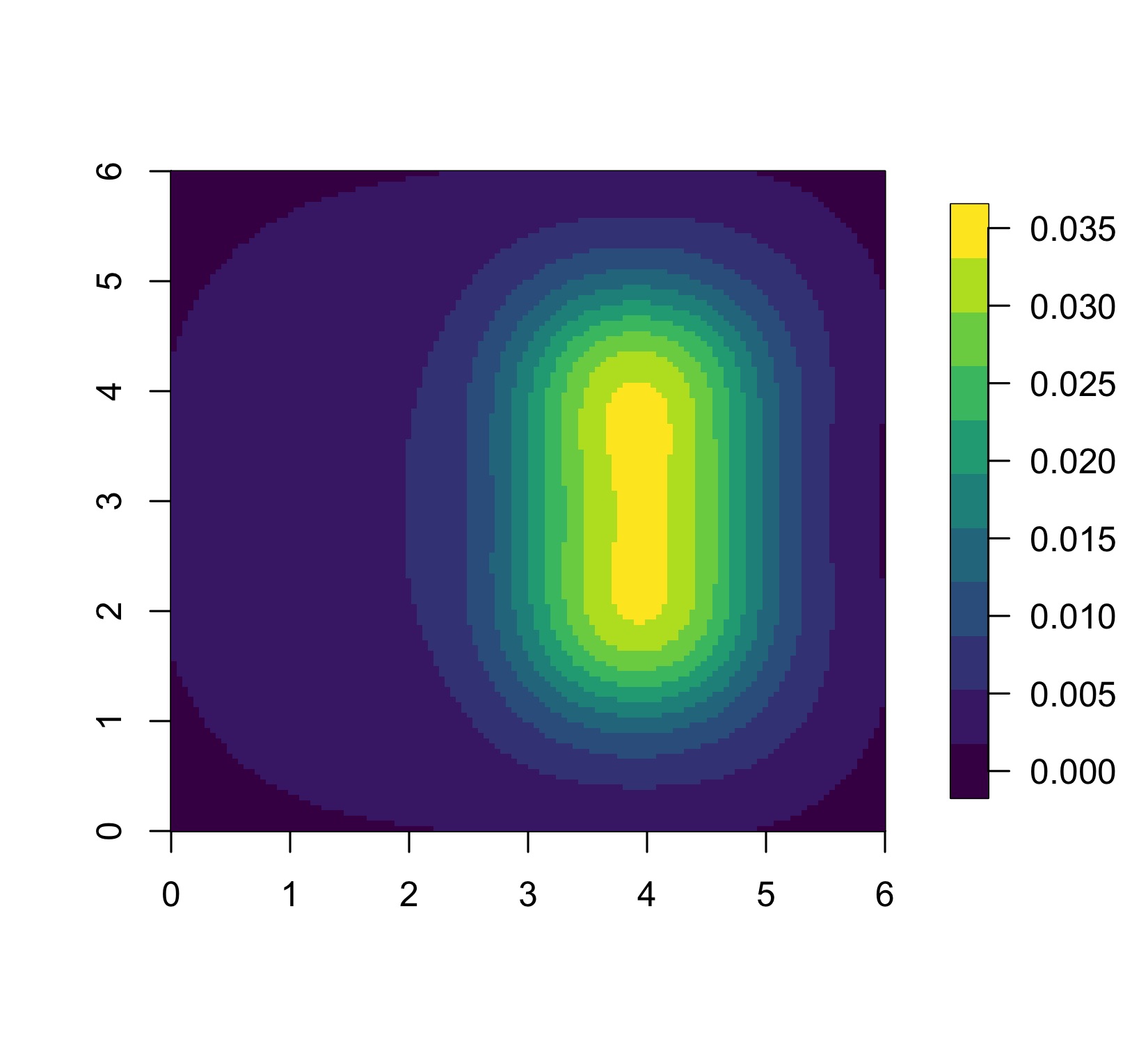}
         \label{fig:synthetic_bgr_mean}
         \caption{CS:1-1, Pixelwise average}
     \end{subfigure}
     \hfill
     \begin{subfigure}[b]{0.49\textwidth}
         \centering
         \includegraphics[width=\textwidth]{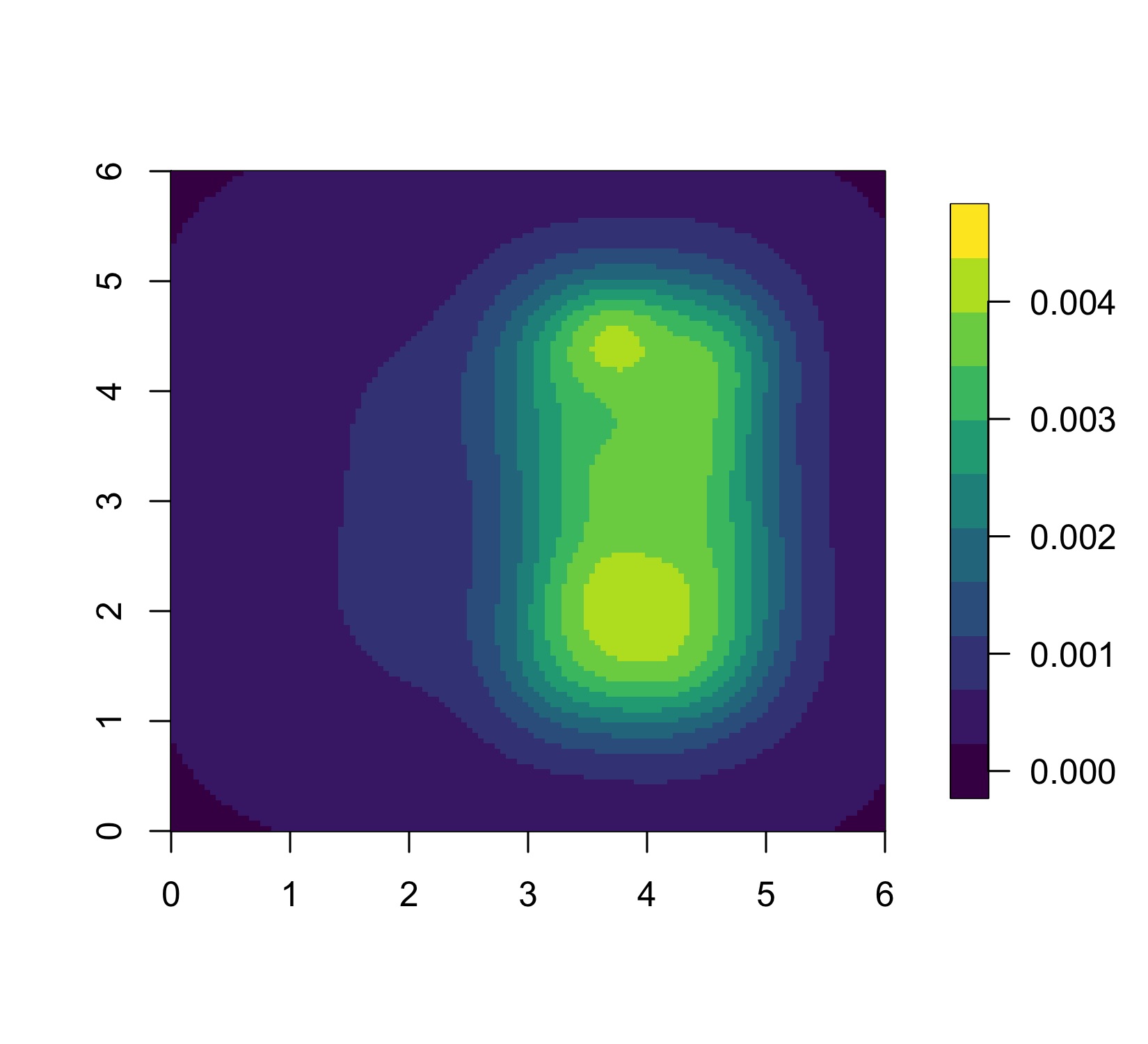}
         \label{fig:synthetic_bgr_sd}
         \caption{CS:1-1, Pixelwise standard deviation}
     \end{subfigure}
     \caption{Estimation results of background rate by the models VS-1:1 and CS-1:1 for 200 synthetic earthquake catalogs.
     }
     \label{fig:synthetic_bgr}
\end{figure}

Now we fit the two models VS-1:1 and CS-1:1 to each of 200 synthetic earthquake catalogs and compare the results. Our result confirms that there is no instability in our results. First, estimated values of $\alpha$ function of the model with spatially varying $\alpha$ are around 1 over the entire spatial domain. Second, background rates are estimated almost identically by both models. Figure \ref{fig:synthetic_bgr} illustrates the pixelwise average and standard deviation of the background rate estimation results by the two models for 200 synthetic earthquake catalogs, and both of them match closely with the assumed background rate.  Lastly, we compare the forecast accuracy of the two models in a same manner as in the section \ref{sec:4}. The average partial AUCs for forecast accuracy are 0.36840 and 0.36836 for both models, respectively, and the pairwise differences of them are very small.

\end{document}